\documentclass[runningheads]{llncs}

\AtBeginDocument{%
  \providecommand\BibTeX{{%
    \normalfont B\kern-0.5em{\scshape i\kern-0.25em b}\kern-0.8em\TeX}}}

\usepackage{multirow}
\usepackage{amssymb}
\usepackage{booktabs}
\usepackage{todonotes}
\usepackage{graphicx}
\usepackage{subcaption} 
\usepackage{float}
\usepackage{adjustbox}
\usepackage{placeins}
\usepackage{url}
\usepackage{breakurl} 
\usepackage{fontawesome5}
\usepackage{enumitem}
\usepackage[most]{tcolorbox}
\usepackage[misc,geometry]{ifsym}

\usepackage{orcidlink}
\usepackage{cite}

\usepackage[normalem]{ulem}
\useunder{\uline}{\ul}{}

\usepackage{pifont}
\newcommand{\cmark}{\ding{51}}%
\newcommand{\xmark}{\ding{55}}%

\begin{document}

\title{To Adopt or Not to Adopt L4S-Compatible Congestion Control? Understanding Performance in a Partial L4S Deployment}
\titlerunning{To Adopt or Not to Adopt L4S-Compatible Congestion Control?}

\author{Fatih Berkay Sarpkaya\textsuperscript{(\Letter)}\orcidlink{0009-0002-3317-9281} \and
Fraida Fund\orcidlink{0000-0002-9897-9282}\and
Shivendra Panwar\orcidlink{0000-0002-9822-6838}}
 
\authorrunning{F. B. Sarpkaya, F. Fund, and S. Panwar}

\institute{NYU Tandon School of Engineering, Brooklyn, NY, USA
\email{\{fbs6417,ffund,panwar\}@nyu.edu}}

\maketitle              

\begin{abstract}

With few exceptions, the path to deployment for any Internet technology requires that there be some benefit to \emph{unilateral} adoption of the new technology. 
In an Internet where the technology is not fully deployed, is an individual better off sticking to the status quo, or adopting the new technology? 
This question is especially relevant in the context of the Low Latency, Low Loss, Scalable Throughput (L4S) architecture, where the full benefit is realized only when compatible protocols (scalable congestion control, accurate ECN, and flow isolation at queues) are adopted at both endpoints of a connection and also at the bottleneck router. 
In this paper, we consider the perspective of the sender of an L4S flow using scalable congestion control, without knowing whether the bottleneck router uses an L4S queue, or whether other flows sharing the bottleneck queue are also using scalable congestion control. 
We show that whether the sender uses TCP Prague or BBRv2 as the scalable congestion control, it cannot be assured that it will not harm or be harmed by another flow sharing the bottleneck link. We further show that the harm is not necessarily mitigated when a scalable flow shares a bottleneck with multiple classic flows. Finally, we evaluate the approach of BBRv3, where scalable congestion control is used only when the path delay is small, with ECN feedback ignored otherwise, and show that it does not solve the coexistence problem.


\end{abstract}

\section{Introduction}~\label{sec:intro}

\vspace{-1em}

High-rate, latency-critical applications, such as online gaming, Virtual Reality (VR), and remote control, are increasingly prevalent on today's Internet. The Low Latency, Low Loss, Scalable Throughput (L4S)~\cite{l4sarch-rfc9330} architecture has attracted a great deal of attention from service providers and equipment vendors~\cite{EricssonDeutscheTelekom2021,comcast2023lowlatency,nokia-hololight,ietf-l4s-deployment-comcast-ietf-121,IETF-120-understanding-prague,livingood-low-latency-deployment-07,WWDC2023-L4S-apple,l4s-interop}, including recent field trials by Comcast~\cite{comcast2023lowlatency}, inclusion in Low Latency DOCSIS~\cite{LowLatencyDOCSIS} specifications, integration in Apple's operating systems~\cite{WWDC2023-L4S-apple}, and adoption in Nvidia's streaming service~\cite{nvidia}. 
L4S is designed to enable both high throughput and low latency while coexisting with classic (not necessarily latency sensitive) flows. It achieves this through scalable congestion control algorithms (e.g., TCP Prague)~\cite{prague-draft-rfc9330,prague-paper}, Accurate Explicit Congestion Notification (AccECN)~\cite{accecn-draft,accecn-rfc7560}, and dual-queue Active Queue Management (AQM)~\cite{dualpi-paper,dualpi-paper2,dualpi-rfc9332}. Unlike other low-latency mechanisms, which may focus solely on specific components, L4S involves both endpoints and routers. When fully deployed, these components enable L4S flows to achieve both high throughput and low latency. However, as traffic traverses the Internet, it encounters network segments with varying levels of L4S support. Some segments may fully implement all L4S components, while others may have partial or no support. As a result, L4S deployment will be incremental, and in the early stages, L4S flows will coexist with other flow types at bottlenecks that may or may be L4S-compatible.

For an endpoint sending data across the Internet, the question of L4S adoption hinges on the decision of whether or not to use a scalable congestion control algorithm (CCA) with AccECN, and a header setting that will classify it into L4S queues at bottleneck routers. This sender must make the decision without full knowledge of whether the bottleneck queue it encounters will be L4S-compatible, or whether other flows sharing the bottleneck will also be scalable flows using AccECN. In this work, we investigate two scalable CCAs, TCP Prague and BBRv2 with AccECN, in partial L4S deployment scenarios including both L4S-compatible and non-L4S bottlenecks, in order to better understand this decision. 

Ultimately, we aim to answer four main research questions:

\newcommand{\RQPrague}{Can senders be assured that TCP Prague will not cause harm to, or be harmed by, another flow at a shared bottleneck?}
\newcommand{\RQBBR}{Does L4S-compatible BBRv2 have more favorable properties for adoption than TCP Prague?}
\newcommand{\RQMulti}{Is the harm caused by or to the L4S-compatible flow mitigated when the bottleneck is shared by a large number of flows? }
\newcommand{\RQBBRThree}{Is the BBRv3 approach more favorable for the deployment of L4S-compatible congestion control?}

\begin{enumerate}
    \item TCP Prague is a scalable congestion control prototype developed by the L4S team. \RQPrague
    
    \item There is an implementation of BBRv2 that adopts some of the principles of scalable congestion control used by TCP Prague. \RQBBR 
    \item \RQMulti
   
    \item In BBRv3, ECN is switched ON or OFF automatically, depending on whether or not a constant path delay threshold is exceeded. \RQBBRThree
\end{enumerate}

Since the sender does not know what type of queue it will encounter and what type of flow will be sharing the bottleneck, if a TCP sender has some reasonable expectation of encountering a pathological situation, then it probably should not adopt scalable congestion control. In our investigation of these research questions, we seek to identify such situations.

This work builds on earlier evaluations of scalable congestion control, both in the academic literature, in the IETF, and industry. The IETF Transport and Services Working Group (TSVWG) is actively working on the development of TCP Prague and the evaluation of L4S. However, beyond some early studies, there has been limited research on L4S evaluation across different scenarios. We will explore this in detail in Section~\ref{sec:Background}. Our previous work~\cite{to-switch-or-not-to-switch-prague} evaluates TCP Prague in specific scenarios and provides some initial findings and conclusions towards Research Question 1 (RQ1). In line with that earlier work and the likely initial deployment of L4S, we consider a residential home broadband scenario, but examine a wider range of CC protocols and bottleneck types that a scalable flow might encounter. This includes two key scalable CCAs, TCP Prague and BBRv2 with AccECN, as well as the most widely used non-scalable CCAs with which a scalable flow may coexist — CUBIC, BBRv1, v2, and v3 — and a greater variety of queue types that might be present at the bottleneck router.

The rest of this paper is organized as follows: Section~\ref{sec:Background} provides a detailed discussion of the L4S architecture, briefly reviews early evaluation results from previous works, and discusses prior work on the deployment of new CCAs and expectations regarding fairness and harm. Section~\ref{sec:methodology} outlines the experimental methodology, including the experiment topology, the CCAs used, and the different bottleneck types. Section~\ref{sec:ExpResults} presents our findings along with a detailed discussion of the results. Finally, Section~\ref{sec:conclusion} concludes with a summary of the work and findings, and suggests directions for future research. Our experimental artifacts are available for use on the open-access testbed FABRIC~\cite{FABRIC}, allowing others to build upon and validate our research\footnote{\label{artifacts}Artifacts are available at: https://github.com/fatihsarpkaya/L4S-PAM2025}. A detailed artifact description is provided in Appendix~\ref{sec:appendix:artifact}.

\vspace{-1em}

\section{Background}\label{sec:Background}

In this paper, we will explore the broader question of whether or not L4S's scalable congestion control is ready for adoption, given that it may be used in a network where L4S is not fully deployed. In order to provide context for this investigation, we provide some background on the L4S architecture (Section~\ref{sec:l4s components}) describe early evaluations of it (Section~\ref{sec:evaluations}), and discuss more generally what it means for a new CCA to be ``ready for adoption'' (Section~\ref{sec:fairness discussion}). 

\vspace{-1em}

\subsection{L4S Architecture}\label{sec:l4s components}

The L4S architecture~\cite{l4sarch-rfc9330} is designed to address the challenge of achieving both high throughput and ultra-low latency. L4S flows should be capable of maintaining very low queuing delay (i.e., less than 1 millisecond), low congestion loss, and high throughput. In order to achieve this, L4S uses three primary mechanisms: scalable congestion control, accurate ECN, and isolation of L4S and non-L4S flows at the bottleneck queue (ideally, with different ECN thresholds).

\begin{figure}[t]
    \centering
    \includegraphics[width=1\textwidth]{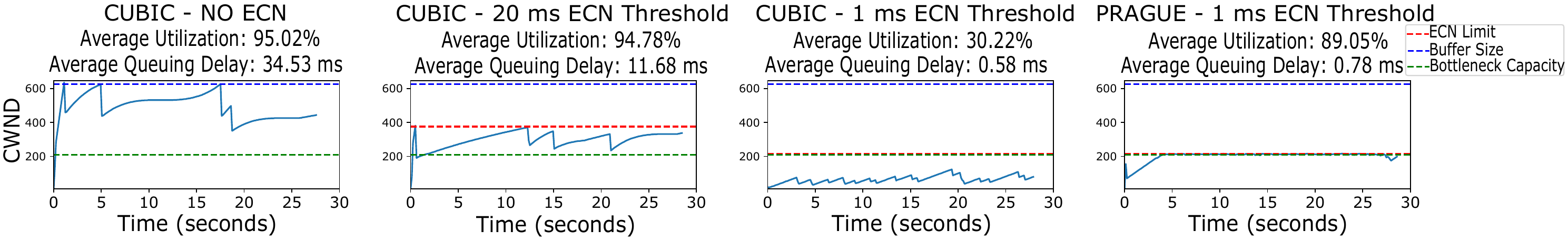}
    \caption{Adapted from \cite{prague-paper} and ~\cite{to-switch-or-not-to-switch-prague}, we conduct a FABRIC experiment to illustrate the fundamental limitation of classic congestion control, using a line network with 100 Mbps bottleneck capacity, a base RTT of 25 ms, and a 2 BDP FIFO bottleneck with ECN. Artifacts to reproduce this are available at~\cite{github}.} 
    \label{fig: Figure1}
    \vspace{-0.7cm}
\end{figure}

\textbf{Scalable Congestion Control} addresses a fundamental problem of classic congestion control, illustrated in Figure~\ref{fig: Figure1}. A sender using a classic loss-based CCA such as TCP CUBIC will increase its congestion window (CWND) until the bottleneck buffer fills, causing high queuing delay (first subplot). Alternatively, ECN marking may be used to signal congestion, potentially reducing the queuing delay somewhat (second subplot). However, due to the sawtooth shape of its CWND - the result of using a fixed (large) multiplicative decrease when it receives a congestion signal - classic congestion control cannot fully utilize the bottleneck link when the ECN marking threshold is set very low, so it cannot achieve extremely low queuing delay (third subplot). Scalable congestion control addresses this by adjusting the CWND in much smaller increments, in proportion to the extent of the congestion, thereby achieving full link utilization with very low queuing delay (fourth subplot). It requires AccECN or DCTP-style ECN marking at the receiver, which we discuss next.

\textbf{DCTCP-style ECN and AccECN.} TCP with classic ECN was designed to indicate at most one congestion event per round trip \cite{ecn-rfc3168}. A classic ECN receiver persistently marks the ECE flag in the TCP header until the sender acknowledges it with a Congestion Window Reduced (CWR) flag. In DCTCP, the receiver provides more precise congestion feedback by ACKing every packet and setting the ECE flag only if the packet has a CE mark. It also accounts for delayed ACKs, using a simple state-machine mechanism to decide whether the ECN-Echo bit should be set or not based on the marking status of the last received packet~\cite{dctcp}. Therefore, the behavior of the receiver is important for senders using the DCTCP-style scalable congestion control mechanism. If the receiver uses classic ECN marking, the sender may not accurately calculate the fraction of CE-marked packets. 
BBRv2 and BBRv3, under some settings, implement scalable congestion control with DCTCP-style ECN.

However, DCTCP-style ECN has some limitations. First, it lacks capability negotiation, assuming both sender and receiver use the same feedback mechanism without verifying compatibility. Second, its feedback becomes unreliable with packet loss, as it cannot effectively distinguish between loss and congestion, causing issues in interpreting feedback, especially with lost ACKs~\cite{accecn-rfc7560}. Due to these issues and the need for more precise feedback in scalable congestion controls within the L4S framework, DCTCP-style ECN is insufficient. 

To address this, AccECN~\cite{ietf-tcpm-accurate-ecn-30} uses a three-bit counter in the TCP header, utilizing the Accurate ECN (AE), CWR, and ECE flags to convey the exact number of packets marked with congestion. This counter, known as the ACE field, provides a precise measure of congestion levels, enabling senders to fine-tune their congestion window with greater precision. Notably, AccECN only governs the signaling of accurate ECN feedback between the TCP sender and receiver, and does not dictate how the sender responds to congestion~\cite{ietf-tcpm-accurate-ecn-30}.

\textbf{Isolation of L4S and non-L4S flows.} Scalable CCAs achieve high throughput and low latency by using AccECN and adjusting their CWND in small increments in proportion to the extent of congestion. However, when these flows share a queue with classic CCAs, one flow type may capture much more of the link capacity than it should. To address this, the L4S architecture introduces the Dual-Queue Coupled AQM, with DualPI2 AQM as a prototype, which separates traffic into two queues: one for low latency flows and another for classic flows. Scalable flows negotiating AccECN should set the ECT(1) code point in the ECN header to be assigned to the low latency queue, while other flows are directed to the classic queue~\cite{dualpi-rfc9332}. This design aims to maintain low latency for scalable flows while ensuring reasonable bandwidth sharing with classic flows. 

In addition to isolating low latency and classic flows, the Dual-Queue Coupled AQM should also use different ECN marking thresholds for the two queues. As illustrated in Figure~\ref{fig: Figure1}, a classic flow cannot tolerate a very low ECN marking threshold, but a scalable flow can.  A queue that provides only flow isolation, such as FQ or FQ-CoDel~\cite{hoeiland2018flow}, can ensure that classic and low latency flows coexist without harm to one another. However, with a common ECN threshold for all queues (presumably, a higher threshold to accommodate the classic flows), the low-latency benefit of L4S cannot be fully realized by L4S flows. The Dual-Queue Coupled AQM and an L4S-aware FQ-CoDel implementation~\cite{fq_codel_l4s} assign a shallower ECN marking threshold for flows that set the ECT(1) code point.

\textbf{What scenarios is an L4S flow likely to encounter?} L4S architecture assumes ideal conditions—scalable congestion control and AccECN support on end hosts, along with flow isolation on routers.
However, actual conditions on the Internet vary widely~\cite{ietf-operational-guidance,heist2021ecn,lim-ecn-traversal}. Some networks deploy ECN as part of AQMs like FQ-CoDel, but others do not. Issues like imperfect flow isolation due to hash collisions, VPN tunneling, or deliberate configurations can also arise. Additionally, misconfiguration, legacy behaviors, and policy decisions (like actively bleaching ECN bits) can unintentionally change or disable ECN functionality. Production traffic observations~\cite{jake-ce-observations,Chun-Xiang-ecn} reveal inconsistent congestion marking capabilities across environments, influenced by equipment age, geographic differences, and network configurations. These studies motivate our evaluations in the next section, considering various network scenarios L4S flows may encounter.

\vspace{-1em}

\subsection{Early Evaluations of L4S Coexistence}\label{sec:evaluations}

A complete L4S deployment involves several mechanisms: scalable CC at the sender, AccECN or DCTCP-style ECN at the receiver, and flow isolation with different ECN thresholds at the bottleneck router. Incremental deployment of these mechanisms, ensuring L4S flows achieve low-latency and high-throughput performance without negatively impacting existing congestion control algorithms, is essential for successful Internet-wide adoption. Consequently, coexistence and deployment strategies have been key subjects of discussion and preliminary evaluation by the IETF TSVWG, industry, and academic literature.

One of the most challenging coexistence scenarios occurs when L4S and Classic CCAs share a single-queue Classic ECN bottleneck. Therefore, White~\cite{ietf-operational-guidance} recommends prioritizing safe coexistence with Classic ECN traffic. It is suggested that the first phase of deployment should focus on enabling L4S-aware AQMs, such as DualQ or FQ-CoDel, at critical points like network edges and bottlenecks~\cite{l4sarch-rfc9330}. This incremental approach allows L4S flows to coexist with Classic flows, providing low-latency and low-loss benefits while minimizing the risks of unsafe coexistence. Deployment typically starts with upgrading AQMs at bottlenecks, followed by introducing L4S-compatible congestion control mechanisms, such as TCP Prague, on servers and end devices. When Classic ECN bottlenecks are detected, additional measures should be taken, like ECN fallback algorithm~\cite{prague-fallback}, to prevent issues like bandwidth starvation for Classic flows~\cite{ietf-operational-guidance}.

However, this deployment strategy has some pre-requisites. First, as discussed, the performance should be validated in various partial deployment scenarios to ensure that L4S can coexist with Classic ECN flows without causing performance degradation. This validation should encourage network operators to upgrade bottleneck routers and adopt scalable CCAs with AccECN feedback at network edges. Another challenge is upgrading bottlenecks, which may occur at locations like peering points or wireless access links, making it critical to focus on these locations~\cite{to-switch-or-not-to-switch-prague}. Additionally, globally upgrading routers presents a significant challenge due to the extensive infrastructure involved. Thus, understanding the behavior of scalable CCAs in different partial L4S deployment scenarios is crucial and will likely remain critical for a long time.

Early evaluations and field trials of L4S have been discussed in IETF meetings~\cite{comcast2023lowlatency,ietf-l4s-deployment-comcast-ietf-121}. \cite{l4s-tests} confirms that with a shared FQ-CoDel single queue or a single PIE queue using classic ECN signaling, TCP Prague flows dominate classic flows, whether or not the classic flows support ECN. \cite{sce-l4s-bakeoff,henderson2019l4s-l4s-testing,henderson2019l4s-issues} also confirm this issue with non-L4S-aware AQMs. Our previous work \cite{to-switch-or-not-to-switch-prague} had similar results for when TCP Prague shares a CoDel queue with CUBIC and non-L4S-compatible BBRv2. 

Notably, \cite{l4s-tests} finds that DualPI2 gives Prague flows a throughput advantage over CUBIC, while in a FIFO queue, CUBIC dominates, with Prague behaving like NewReno. These observations are confirmed in~\cite{henderson2019l4s-l4s-testing}. However, our previous work \cite{to-switch-or-not-to-switch-prague} shows that CUBIC or non-L4S-compatible BBRv2 flows gain a slight advantage over Prague flows in DualPI2, and in a FIFO queue, Prague and CUBIC share bandwidth almost equally. This discrepancy is possibly due to differences in network setups, which can affect fairness as discussed in \cite{sce-l4s-bakeoff,boruoljira2020validating}.

Most recently, we had observed~\cite{to-switch-or-not-to-switch-prague} that across various bottleneck types and co-existence scenarios involving TCP Prague versus CUBIC or non-L4S-compatible BBRv2, only bottlenecks with per-flow isolation ensure a fair share of bandwidth, while only DualPI2 guarantees ultra-low latency for TCP Prague. However, this paper does not consider the full range of queue types or scalable CCAs that may be involved in an incremental L4S deployment.

\vspace{-1em}

\subsection{``Ready for adoption''}\label{sec:fairness discussion}

Beyond L4S specifically, there is a rich academic literature on what it means for a CCA to be ready for adoption, that provides important context for our study.

The classic metric used to assess CCA fairness is Jain's Fairness Index (JFI)~\cite{jain's-fairness}. However, several studies explore broader TCP fairness concepts and propose alternative metrics. In \cite{prudentia-mmf-fairness}, fairness is evaluated using max-min fairness (MmF), focusing on how closely each flow achieves its MmF share during contention. This approach measures how much of the allocated bandwidth each flow gets relative to its fair share. In \cite{cloud-based-fairness}, fairness is calculated by normalizing the throughput difference between competing flows, with 0 representing perfect fairness, and deviations indicating one flow taking more bandwidth than the other. Similarly, \cite{congestion-control-wild} introduces a ratio-based fairness measure, which evaluates fairness by comparing the bytes transmitted between two flows. A value of 0 indicates perfect fairness, while negative or positive values indicate one flow dominates the other. 

In \cite{ware-jain-fairness}, the authors argue that perfect flow rate fairness is not always necessary and that JFI alone does not capture positive or negative biases between flows. They propose that new CCAs should avoid causing harm to existing algorithms and introduce a harm-based threshold to measure a new CCA's negative impact on legacy algorithms. The goal is to ensure a new algorithm does no more harm than existing algorithms do to each other. This harm-based metric is used in \cite{bbr-vs-bbr2-harm-metric-fairness}, for example, to evaluate the fairness between BBR and BBRv2, showing how newer algorithms can be assessed through their impact on legacy flows. Furthermore, \cite{is-it-necessary-beyond-jain-fairness} confirms that the harm-based approach better evaluates CCA deployability in real-world networks compared to traditional fairness metrics.

Additionally, as discussed in \cite{rfc5033bis}, new CCAs should be evaluated against existing algorithms, ensuring that any new proposal does not introduce more harm than previous standards-track algorithms to flows sharing a common bottleneck. This approach ensures that newer CCAs do not unfairly dominate bandwidth at the expense of legacy flows, which is essential for incremental deployment.

From the discussion above it is clear that there are various methods to evaluate the extent to which a CCA achieves fairness, and no single approach is universally best. Therefore, our fairness evaluation reflects the principles discussed in \cite{rfc5033bis} and \cite{ware-jain-fairness}, emphasizing harm-based fairness rather than strict 50-50 
 equality, which is not necessarily expected.
Our evaluations aim to inform senders that ``if you use scalable congestion control in this type of network, you may experience this behavior.'' The sender can then decide what best suits their application’s needs and whether switching to scalable congestion control is worthwhile. For this reason, we will report throughput directly (instead of JFI) in Section~\ref{sec:ExpResults}. 
However, although we report results for all scenarios, in our discussion we focus especially on those where a flow gets less than 20\% of bottleneck link bandwidth or more than 80\% (when it should get 50\%). This heuristic helps us identify the most problematic settings. 

\vspace{-1em}

\section{Experiment Methodology}\label{sec:methodology}

\vspace{-1em}

To address the research questions in Section~\ref{sec:intro}, we conduct systematic experiments and report the results. This section details our methodology.

\textbf{Experiment Platform:} We conduct experiments on FABRIC \cite{FABRIC}, a national scale experimental networking testbed. Each node in our experiment is a virtual machine running Ubuntu 22.04, with 4 cores and 32 GB RAM for single flow experiments and 8 cores and 64 GB RAM for multiple flow experiments.

\begin{figure}
    \centering
    \includegraphics[width=0.5\textwidth]{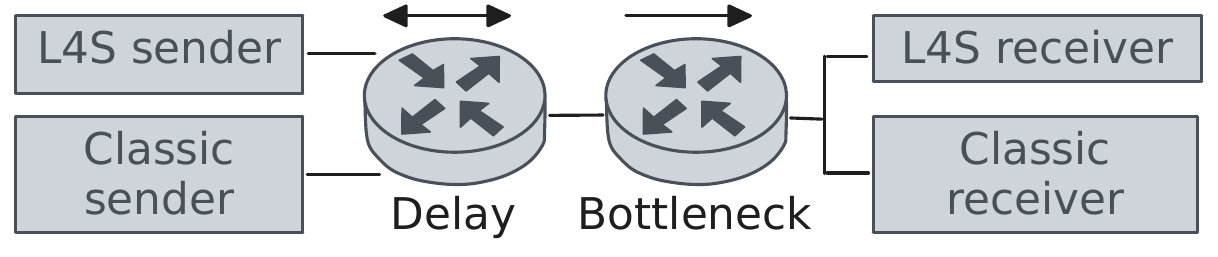}
    \caption{Experiment Topology}
    \label{fig: Figure2}
    \vspace{-0.8cm}
\end{figure}

\textbf{Topology:} Our network setup consists of two senders and two receivers, along with a delay node and a bottleneck router, as shown in Figure~\ref{fig: Figure2}. The delay node uses \texttt{netem} to emulate a base RTT, with half of the delay applied in each direction. At the router, we configure the bottleneck bandwidth and buffer size using the token bucket filter implemented in \texttt{tc-htb}.

\textbf{Network Setting:} We assume a bottleneck at the access link and emulate network conditions similar to those in a residential fixed broadband scenario, with a base RTT of 10 ms and a bottleneck link capacity of 100 Mbps. We choose to study the residential broadband setting in depth because of L4S's early momentum among cable providers~\cite{LowLatencyDOCSIS,comcast2023lowlatency,ietf-l4s-deployment-comcast-ietf-121}.

\textbf{Queue:} In our experiments, we consider a wide variety of queue types that may be encountered at the bottleneck router.

First, we consider a variety of single queue types, including three that drop packets as a congestion signal and two that use ECN to mark packets:

\begin{itemize}[noitemsep, topsep=0pt]
  \item \textbf{FIFO:} a single drop tail queue without ECN support, realized with \texttt{tc-bfifo}.

  \item \textbf{Drop-based CoDel:} a single queue with CoDel AQM~\cite{codel}, which uses the local minimum queue size within a monitoring window as a measure of the standing queue, and drop packets if there is a standing queue exceeding a target value. We use \texttt{tc-codel} with a 5~ms target and ECN disabled.

  \item \textbf{Drop-based PIE:} a single queue with PIE~\cite{rfc8033-pie} estimates the dequeue rate to calculate the current queue delay, adjusting the drop probability based on delay trends. We use \texttt{tc-pie} with a 5~ms target and ECN disabled.

  \item \textbf{FIFO + ECN:} a single drop tail queue with ECN support using a 5~ms marking threshold, realized with \texttt{tc-fq}. (Although \texttt{tc-fq} is multi-queue, we enforce it to operate as a single queue.) While such a bottleneck type is considered rare or non-existent in the Internet~\cite{ietf-operational-guidance}, it allows us to distinguish the effect of ECN from the effect of AQM, providing better insight into the mechanisms behind the results.
  
  \item \textbf{CoDel:} ECN enabled version of Drop-based CoDel AQM. We use \texttt{tc-codel} with a 5~ms target and the ECN option enabled, so that it marks packets for flows with ECN support and drops packets otherwise.
\end{itemize}

We also consider some queue types that implement multiple queues. All of these offer flow isolation, but the two queue types marked with a $\star$ also have separate ECN marking thresholds for realizing the full benefits of L4S.

\begin{itemize}[noitemsep, topsep=0pt]
  
  \item \textbf{FQ:} a fair queue with flow isolation and ECN (using a 5~ms marking threshold), realized with \texttt{tc-fq}. 

  \item \textbf{FQ-CoDel:} combines fair queuing with the CoDel AQM. We realize this queue with \texttt{tc fq\_codel}, with a 5~ms target and the ECN option enabled. 

  \item $\star$ \textbf{L4S-aware FQ-CoDel:} extends the traditional FQ-CoDel algorithm to support L4S, allowing a shallow queueing threshold to be applied specifically to L4S packets. We realize this queue using \texttt{tc fq\_codel}, with a 5~ms target and ECN enabled. Additionally, the \texttt{ce\_threshold} is set to 1~ms, and \texttt{ce\_threshold\_selector} is configured as 0x01/0x01 to apply the low threshold exclusively to L4S flows.

  \item $\star$ \textbf{DualPI2:} a dual queue coupled AQM designed for L4S \cite{dualpi-rfc9332}, realized using \texttt{tc-dualpi2} from the L4S repository \cite{l4srepo} (commit \texttt{4579ffb}). We use the \texttt{target} parameter to set a 5~ms ECN threshold for the classic queue, while the L4S queue has a 1 ms ECN threshold.

\end{itemize}

For each queue type, we evaluate bottleneck buffer sizes that range from shallow to deep, considering the following multiples of the link's bandwidth-delay product (BDP): 0.5, 1, 2, 4, and 8.

\textbf{Flow Generation}: We generate TCP flow(s) from the L4S sender and from the classic sender using the \texttt{iperf3} tool, each for a duration of 60 seconds. For each flow, we record the average throughput and RTT values. The results presented are based on the averages from 10 trials.

\textbf{L4S Flow CCA:} There are widely available Linux implementations for two TCP CCAs - Prague and BBRv2 - that implement both scalable congestion control and ECT(1) code point  setting (in order to be assigned to the low latency queue of an L4S bottleneck). Both of these use AccECN. We use the implementation from the L4S repository \cite{l4srepo} (commit \texttt{4579ffb}). Notably, this L4S-compatible BBRv2 is not the ``official'' BBRv2 (that uses DCTCP-style ECN marking, and does not use the ECT(1) code point setting).

The Prague implementation includes an ECN Fallback heuristic~\cite{prague-fallback}, an optional feature to address fairness issues between TCP Prague and classic CCAs when they share a single queue with classic ECN. It allows Prague to fallback to classic behavior if it detects a classic ECN in use. While this feature is not enabled by default in the Linux implementation of Prague, it can be activated. We evaluate Prague both with the default setting, and with ECN Fallback activated.

\textbf{Non-L4S Flow CCA:} For the non-L4S flows, we focus on two CCAs that L4S flows are likely to encounter at a shared bottleneck. 

TCP CUBIC~\cite{cubic} is the most widely used TCP variant on the Internet~\cite{census,cubic-prevalent-ietf-rfc-9438}. We experiment with a classic CUBIC flow (using the Linux kernel 5.13.12 implementation), with and without ECN enabled at both endpoints.

TCP BBR~\cite{cardwell2016bbr} and its variants are widely deployed across the Internet~\cite{census}, are used for all of Google's internal WAN traffic, and for public Internet traffic from services like Google.com and YouTube~\cite{bbrv3-ietf-120-slides}. With respect to ECN, we consider a wide variety of ECN settings: Our experiments include BBRv1 (Linux kernel 5.13.12), which does not support ECN. The default BBRv2 implementation (from the \texttt{v2alpha} branch of the official BBR repository~\cite{bbrrepo}) is used without ECN, with DCTCP-style ECN at the sender and classic ECN at the receiver (one-sided DCTCP-style ECN), or with DCTCP-style ECN at both endpoints (two-sided DCTCP-style ECN). 

We also consider BBRv3 using the \texttt{v3} branch of the official repository~\cite{bbrrepo}.  While BBRv3 has been recently released, the developers do not consider it ready for widespread deployment with its current coexistence and loss-handling properties~\cite{bbr-discussion-google-groups}. Nevertheless, we are interested in exploring its potential. For BBRv3, even if ECN is enabled, it is not utilized when the path delay exceeds a fixed threshold. We explore this behavior in Section~\ref{subsec:EvalBBRv3}.


\section{Experiment Results}\label{sec:ExpResults}

In this section, we discuss the performance of L4S-compatible congestion controls — TCP Prague and L4S-compatible BBRv2 — focusing on their throughput share and latency across various network scenarios, as previously described. We will also address our research questions throughout the discussion.

\vspace{-1em}

\subsection{RQ1: \RQPrague}\label{sec:RQ1}

To address our first research question, we evaluate the performance of TCP Prague and assess whether it is negatively impacted by other flows sharing the same bottleneck, or if it causes harm to those other flows.

\begin{figure*}

  \centering
  \includegraphics[width=1.0\textwidth]{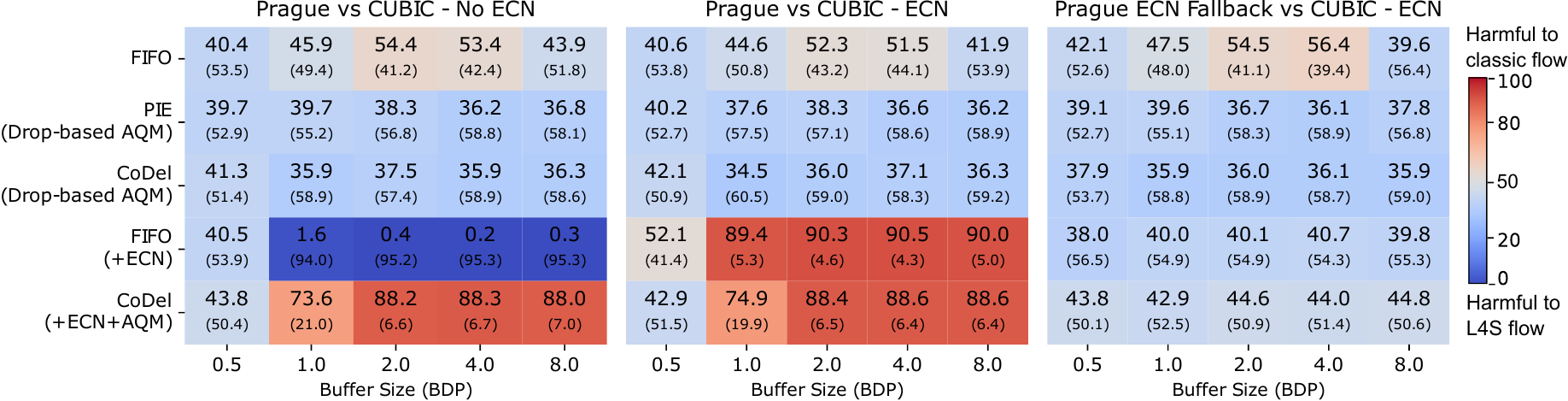}
  \caption{Throughput of Prague (and CUBIC, in parentheses) in Mbps when sharing a bottleneck (Single Queue AQMs).}
  \label{fig:Prague Throughput vs Cubic - Single Queue}
  \vspace{0.5cm}

 \centering
  \includegraphics[width=1.0\textwidth]{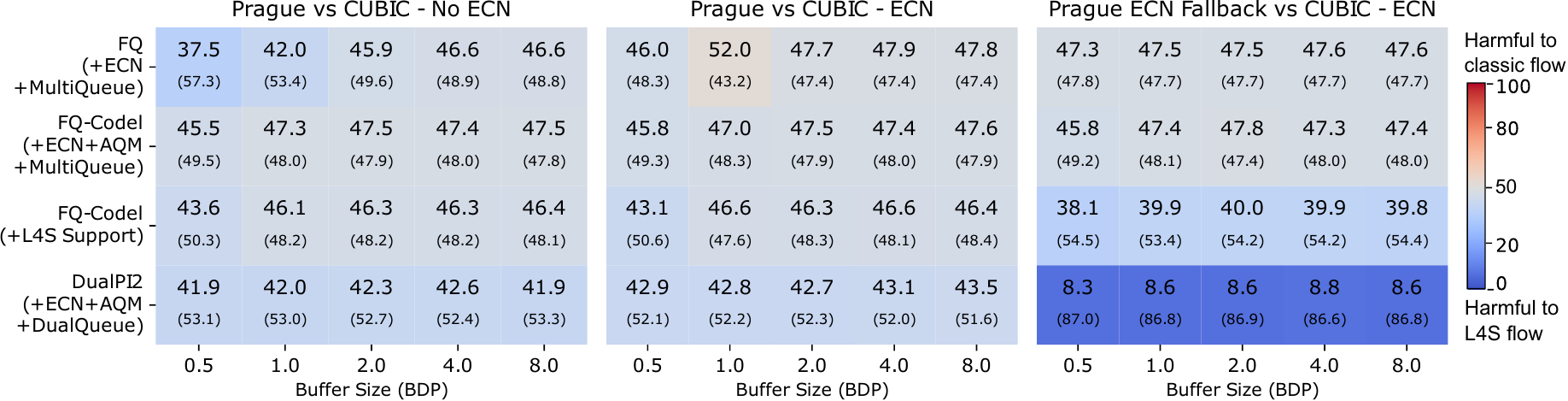}
  \caption{Throughput of Prague (and CUBIC, in parentheses) in Mbps when sharing a bottleneck (multi/dual queue AQMs).}
  \label{fig:Prague Throughput vs Cubic - Multi Queue}
  \vspace{0.5cm}

 \centering
  \includegraphics[width=1.0\textwidth]{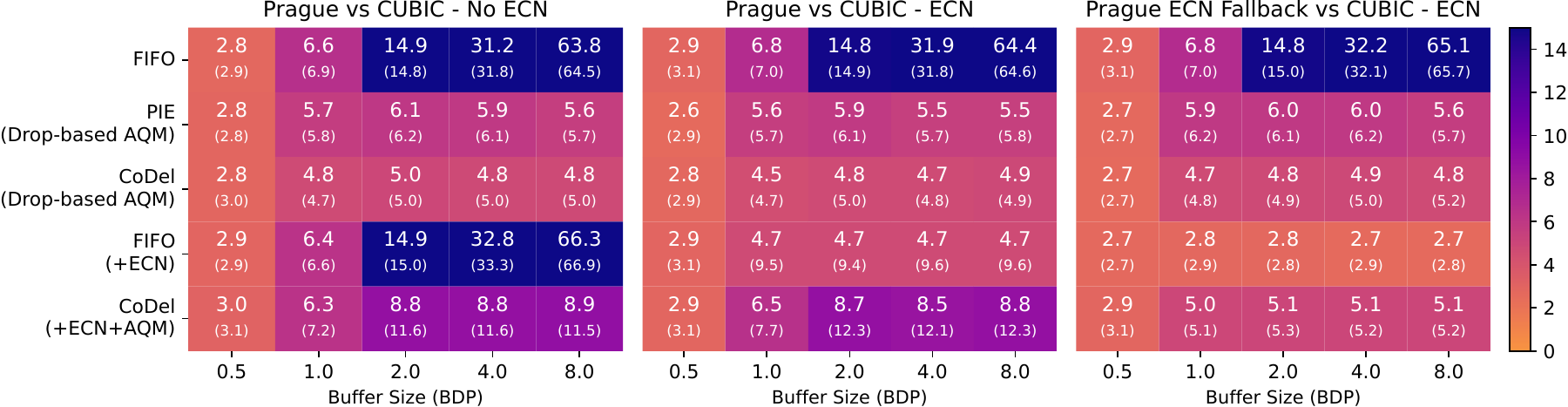}
  \caption{Prague (CUBIC) queuing delay in ms when sharing bottleneck with a CUBIC flow. (single queue AQMs - ECN threshold is 5 ms, where applicable.)}
  \label{fig: Prague Latency vs Cubic - Single Queue}
  \vspace{0.5cm}

\centering
  \includegraphics[width=1.0\textwidth]{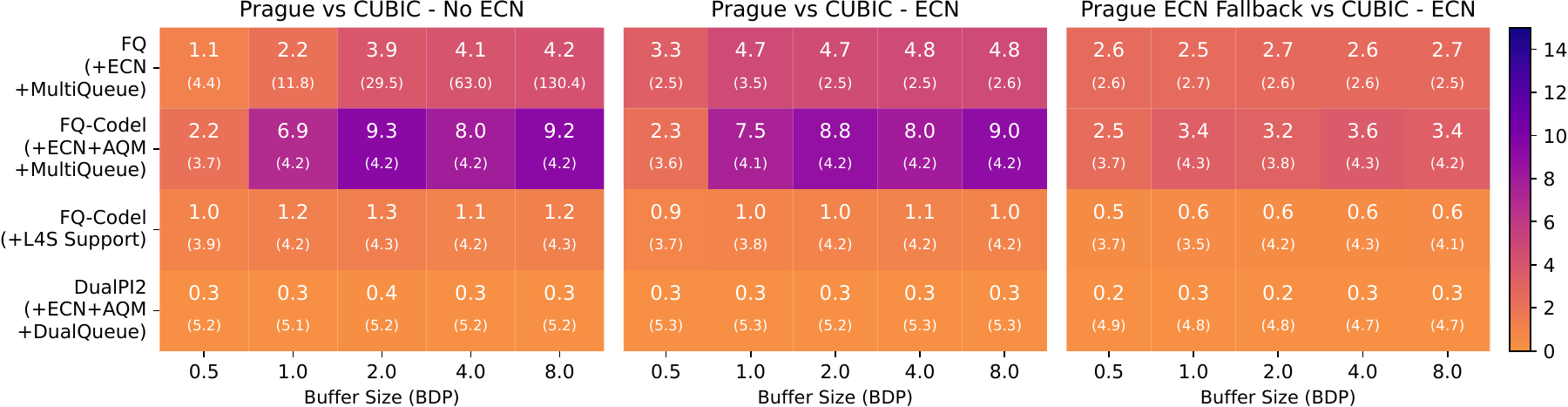}
  \caption{Prague (and CUBIC, in parentheses) queuing delay (ms) when sharing a bottleneck (multi/dual queue AQMs).}
  \label{fig: Prague Latency vs Cubic - Multi Queue}
\end{figure*}

First, we will evaluate the throughput and latency of a TCP Prague flow when it shares a bottleneck queue with a CUBIC flow. The throughput of the Prague flow (and the CUBIC flow, in parenthesis) is shown in Figure~\ref{fig:Prague Throughput vs Cubic - Single Queue} for single queue bottleneck scenarios, and in Figure~\ref{fig:Prague Throughput vs Cubic - Multi Queue} for multi-queue bottleneck scenarios. The latency for each experiment is shown in Figure~\ref{fig: Prague Latency vs Cubic - Single Queue} and Figure~\ref{fig: Prague Latency vs Cubic - Multi Queue}.

\textit{Circumstances where Prague is not harmed by, nor harmful to, a CUBIC flow.} These circumstances include drop-based bottlenecks, FQ bottlenecks, and DualPI2. Regarding the drop-based bottlenecks, as shown in Figure~\ref{fig:Prague Throughput vs Cubic - Single Queue}, in FIFO queues without ECN support or AQM, TCP Prague achieves roughly its fair share, with slight advantages or disadvantages depending on the buffer size. Similarly, in drop-based bottleneck types that use AQMs such as PIE and CoDel, Prague shows slightly worse performance compared to CUBIC (Figure~\ref{fig:Prague Throughput vs Cubic - Single Queue}). In these cases, since ECN is inactive, the only congestion signal is the packet loss. When a loss is detected, Prague falls back to Reno behavior in the current Linux implementation~\cite{briscoe-iccrg-prague-congestion-control-04}. It is well-established that CUBIC generally achieves higher throughput than Reno when sharing the same bottleneck~\cite{cubic-vs-reno}. Therefore, Prague, which effectively behaves like Reno, gets slightly lower throughput in most of the cases in our setting. 

At FQ bottlenecks, as seen in Figure~\ref{fig:Prague Throughput vs Cubic - Multi Queue}, with or without ECN support of CUBIC flow, Prague gets almost its fair share in all scenarios. In L4S-aware FQ-CoDel, Prague flows also get very low queuing delay of around 1ms (Figure~\ref{fig: Prague Latency vs Cubic - Multi Queue}). In DualPI2, Prague similarly gets almost its fair share with a slight disadvantage compared to CUBIC and its queueing delay is also lower than 0.5ms.

\textit{Circumstances where Prague is harmed due to coexistence with a CUBIC flow.} In FIFO with ECN support, there are no AQM mechanisms; it is simply a drop-tail queue with ECN marking capability. In this type of queue, Prague experiences dramatically reduced throughput when competing with CUBIC without ECN (Figure~\ref{fig:Prague Throughput vs Cubic - Single Queue}), since there is no extra mechanism to limit the non-ECN flow. Moreover, in this bottleneck, Prague's queueing latency is extremely high, especially in deep buffers (Figure~\ref{fig: Prague Latency vs Cubic - Single Queue}). This shows that there are scenarios where Prague still experiences extremely high latency, even if it enables its ECN and goes through an ECN-enabled bottleneck.
Prague with ECN fallback, which we discuss next, is also harmed by coexistence with CUBIC in a DualPI2 bottleneck.

\textit{Circumstances where Prague harms a CUBIC flow.} In single queue with ECN bottlenecks, Prague can be harmful. When competing with CUBIC that supports ECN, Prague harms the CUBIC flow due to their different reactions to ECN signals (Figure~\ref{fig:Prague Throughput vs Cubic - Single Queue}). This is a well-known issue, as discussed in Section~\ref{sec:Background}. 
In CoDel AQM with ECN support (Figure~\ref{fig:Prague Throughput vs Cubic - Single Queue}), CoDel AQM drops non-ECN packets and marks ECN-capable packets, allowing Prague to maintain higher throughput. 

This dominance of TCP Prague appears to be mitigated for this type of queue by Prague's ECN fallback algorithm. Prague detects the single-queue AQM with classic ECN support and falls back to classic ECN behavior. In this case, it achieves slightly lower throughput compared to its fair share, similar to what is observed in drop-based AQM scenarios (Figure~\ref{fig:Prague Throughput vs Cubic - Single Queue}). However, it introduces other problems with DualPI2, as seen in Figure~\ref{fig:Prague Throughput vs Cubic - Multi Queue}, where Prague with ECN Fallback gets much less than its share of the link capacity.
This is likely due to the heuristic incorrectly detecting that the queue is not using L4S AQM, as we suggested in~\cite{to-switch-or-not-to-switch-prague}.

\begin{figure*}

  \centering
  \includegraphics[width=1.0\textwidth]{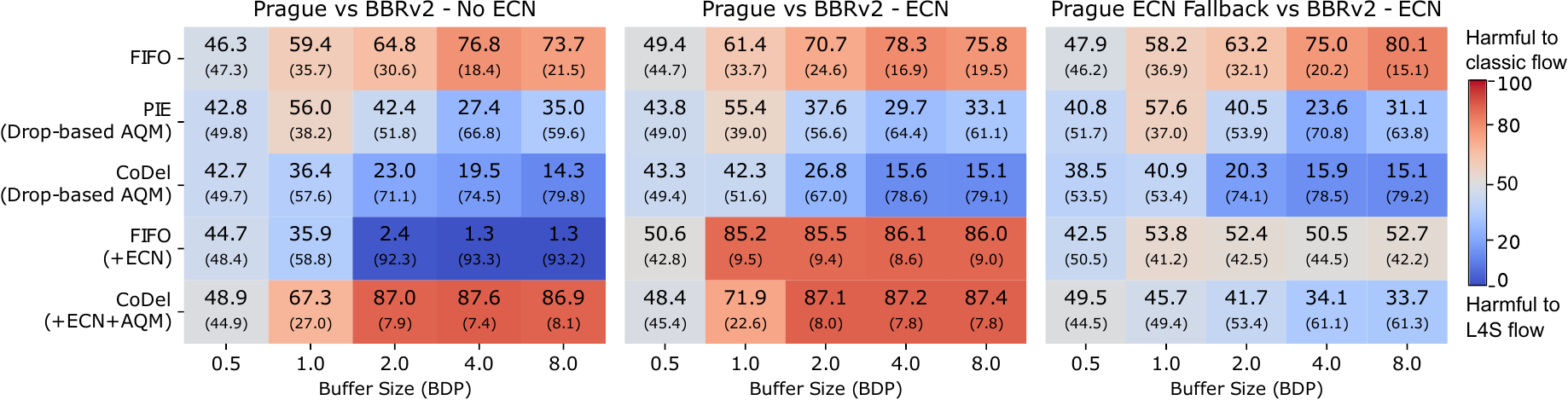}
  \caption{Throughput of Prague (and BBRv2, in parentheses) in Mbps when sharing a bottleneck (single queue AQMs). (classic ECN marking on BBRv2 receiver)}
  \label{fig:Prague Throughput vs BBRv2 - Single Queue}
  \vspace{0.5cm}

  \centering
  \includegraphics[width=1.0\textwidth]{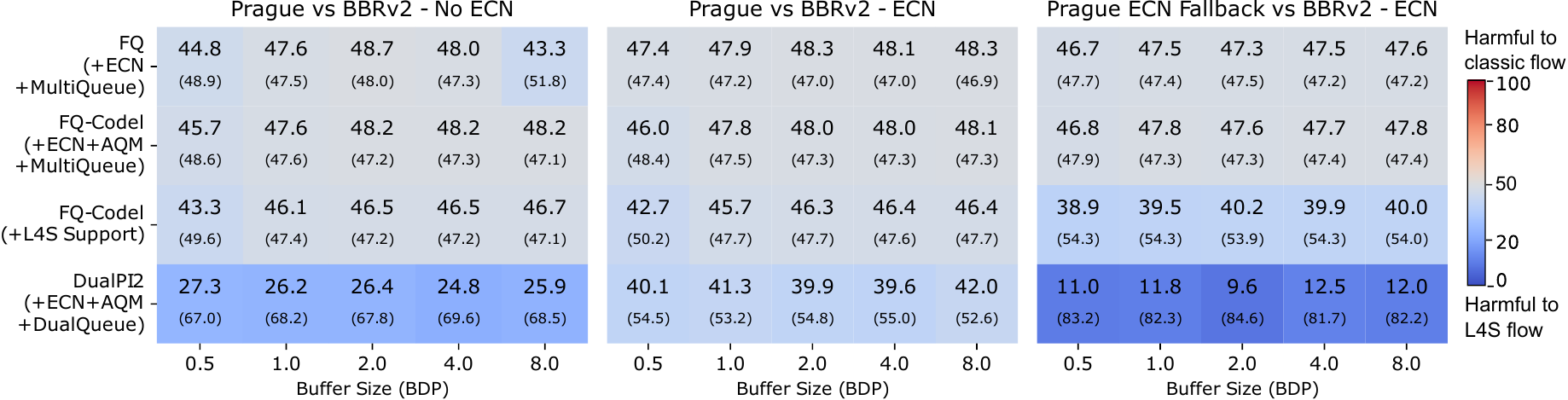}
  \caption{Throughput of Prague (and BBRv2, in parentheses) in Mbps when sharing a bottleneck (multi/dual queue AQMs). (classic ECN marking on BBRv2 receiver)}
  \label{fig:Prague Throughput vs BBRv2 - Multi Queue}
  \vspace{0.5cm}

  \centering
  \includegraphics[width=1.0\textwidth]{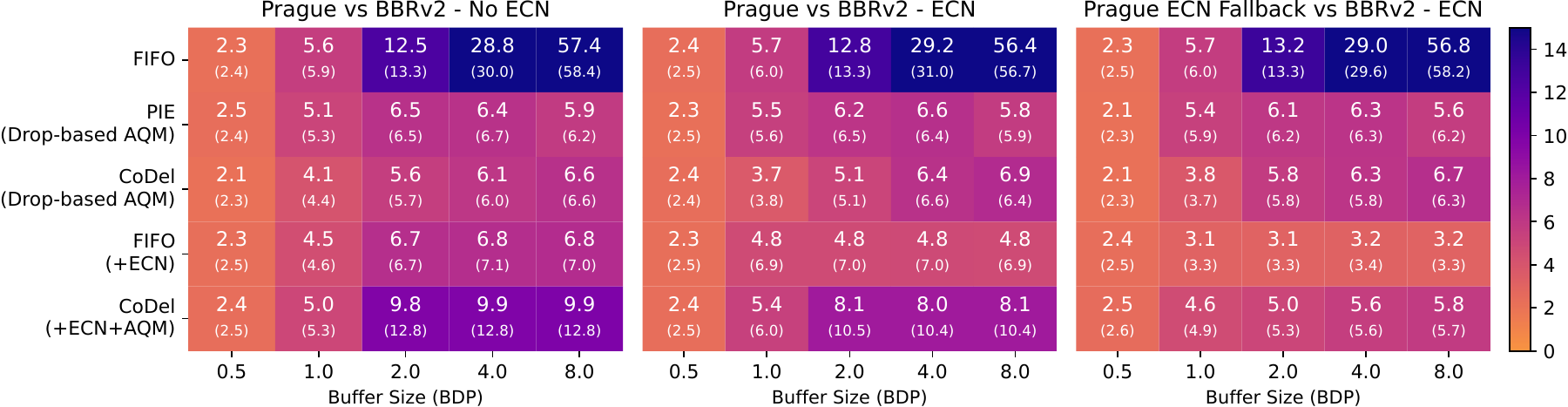}
  \caption{Prague (and BBRv2, in parentheses) queuing delay (ms) when sharing a bottleneck (single queue AQMs). (classic ECN marking on BBRv2 receiver)}
  \label{fig: Prague Latency vs BBRv2 - Single Queue}
  \vspace{0.5cm}

  \centering
  \includegraphics[width=1.0\textwidth]{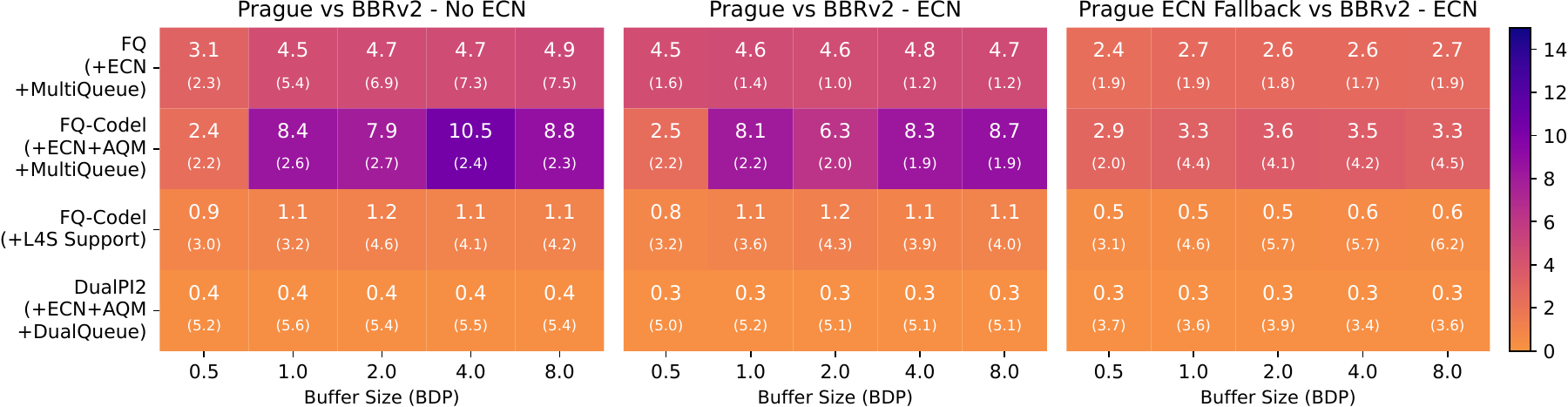}
  \caption{Prague (and BBRv2, in parentheses) queuing delay (ms) when sharing a bottleneck (multi/dual queue AQMs). (classic ECN marking on BBRv2 receiver)}
  \label{fig: Prague Latency vs BBRv2 - Multi Queue}
\end{figure*}

\begin{figure*}

        \centering
        \includegraphics[width=1.0\textwidth]{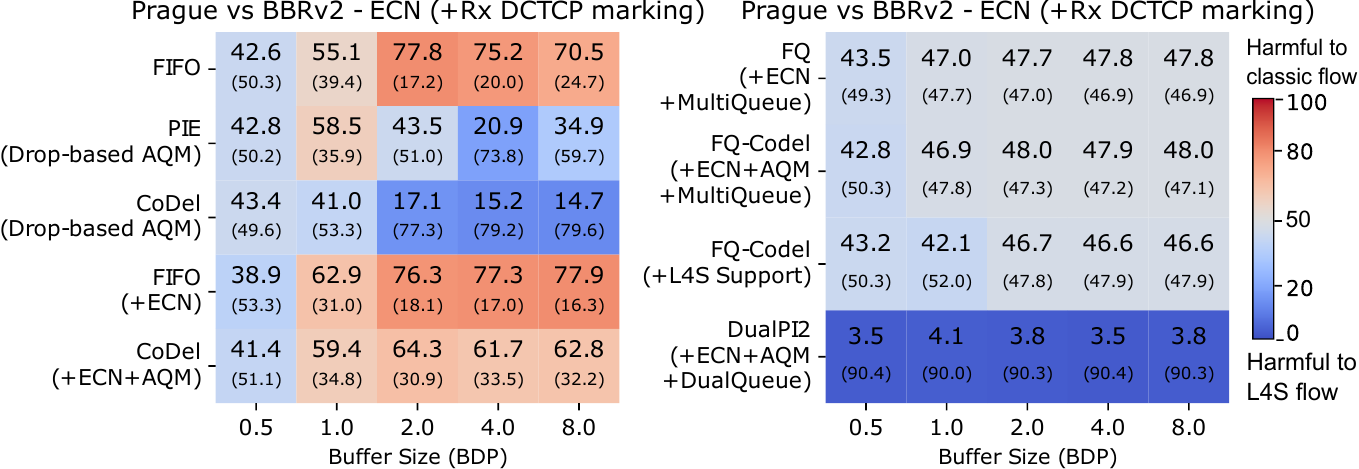}
        \caption{Throughput of Prague (and BBRv2, in parentheses) in Mbps when sharing a bottleneck. (DCTCP-style ECN marking on BBRv2 receiver)}
        \label{fig:tput-bbr2-dctcp}

        \centering
        \includegraphics[width=1.0\textwidth]{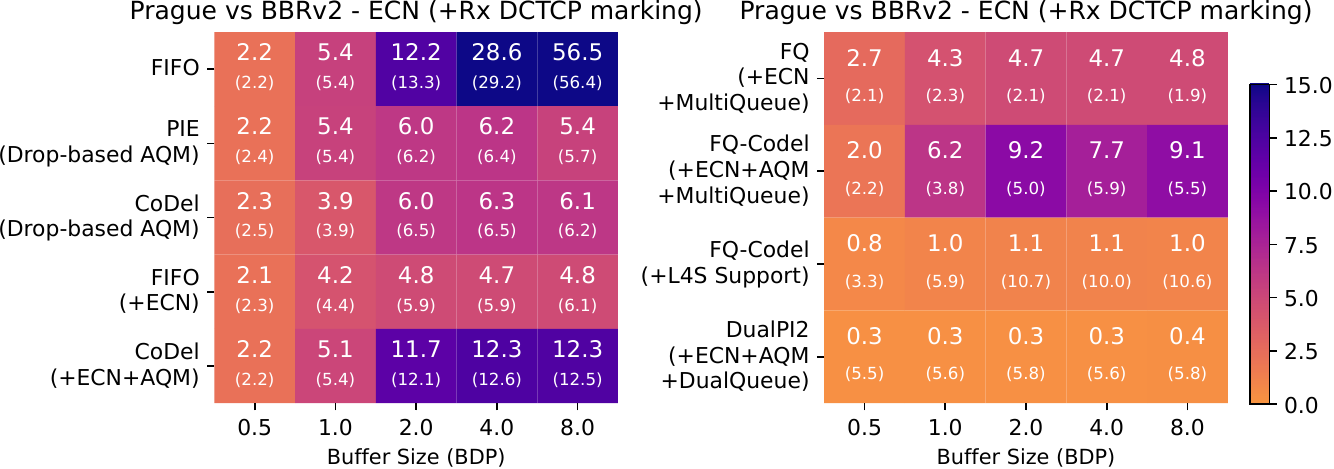}
        \caption{Prague (and BBRv2, in parentheses) queuing delay (ms) when sharing a bottleneck. (DCTCP-style ECN marking on BBRv2 receiver)}
        \label{fig:latency-bbr2-dctcp}

        \centering
        \includegraphics[width=1.0\textwidth]{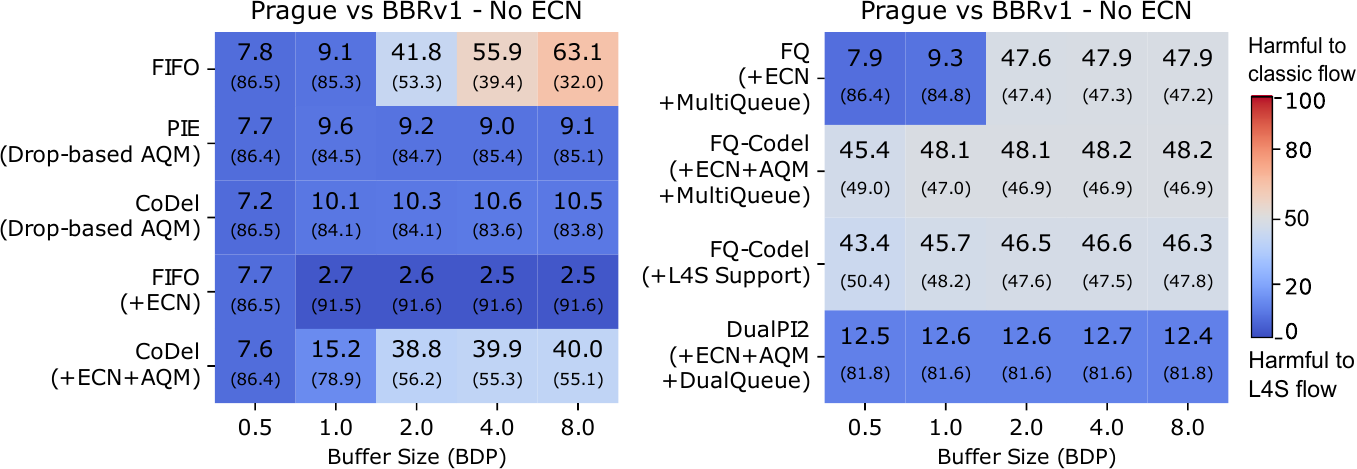}
        \caption{Throughput of Prague (and BBRv1, in parentheses) in Mbps when sharing a bottleneck.}
        \label{fig:tput-bbr1}

        \centering
        \includegraphics[width=1.0\textwidth]{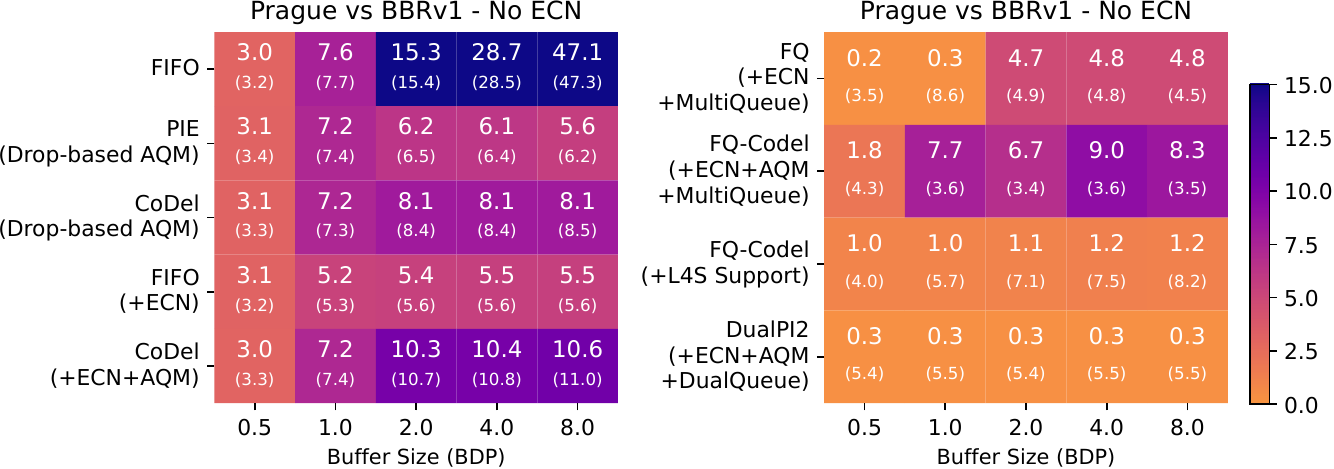}
        \caption{Prague (and BBRv1, in parentheses) queuing delay (ms) when sharing a bottleneck.}
        \label{fig:latency-bbr1}
\end{figure*}

Next, we evaluate TCP Prague against BBRv1/v2 flows. Since BBR is the second-most prevalent TCP variant among major websites~\cite{census}, it is likely that TCP Prague flows may need to coexist with BBR on the Internet. For BBRv2, we consider two different ECN configurations: one-sided DCTCP-style ECN support (i.e. the receiver sends classic ECN ACKs), and two-sided DCTCP-style ECN support (i.e. the receiver sends DCTCP style ACKs).  For coexistence scenarios with BBRv2, the throughput of the Prague flow (and the BBR flow, in parenthesis) is shown in Figure~\ref{fig:Prague Throughput vs BBRv2 - Single Queue} for single queue bottlenecks, and in Figure~\ref{fig:Prague Throughput vs BBRv2 - Multi Queue} for multi-queue bottlenecks. The latency for each experiment is shown in Figure~\ref{fig: Prague Latency vs BBRv2 - Single Queue} and Figure~\ref{fig: Prague Latency vs BBRv2 - Multi Queue}. We also show results vs. BBRv2 with DCTCP style ECN ACKs from the BBRv2 receiver in Figure~\ref{fig:tput-bbr2-dctcp} and Figure~\ref{fig:latency-bbr2-dctcp}. Finally, in Figure~\ref{fig:tput-bbr1} and Figure~\ref{fig:latency-bbr1} we report results for TCP Prague coexistence with BBRv1.

\textit{Circumstances where Prague is not harmed by, nor harmful to, a BBR flow.} FQ bottlenecks are mostly successful to provide fairness for both BBRv1 (Figure~\ref{fig:tput-bbr1}) and BBRv2 (Figure~\ref{fig:Prague Throughput vs BBRv2 - Multi Queue} and~\ref{fig:tput-bbr2-dctcp}) when they share the bottleneck with TCP Prague flows. In shallow buffer single queue drop-based bottlenecks with BBRv2, Prague gets close to its fair share with a slight advantage or disadvantage (Figure~\ref{fig:Prague Throughput vs BBRv2 - Single Queue} and~\ref{fig:tput-bbr2-dctcp}). In FIFO with large buffer sizes, Prague gets a slight advantage compared to BBRv1 (Figure~\ref{fig:tput-bbr1}) and BBRv2 (Figure~\ref{fig:Prague Throughput vs BBRv2 - Single Queue} and~\ref{fig:tput-bbr2-dctcp}). In DualPI2, Prague faces a slight disadvantage against BBRv2 as depicted in Figure~\ref{fig:Prague Throughput vs BBRv2 - Multi Queue}. The fallback algorithm for Prague works well and resolves co-existence issues with BBRv2 in single-queue ECN bottlenecks (Figure~\ref{fig:Prague Throughput vs BBRv2 - Single Queue}). However, it introduces other problems with DualPI2 similar to CUBIC co-existence, which will be discussed later. In terms of latency, L4S-aware AQMs are successful in providing ultra-low latency for Prague competing with BBRv1 (Figure~\ref{fig:latency-bbr1}) and BBRv2 (Figure~\ref{fig: Prague Latency vs BBRv2 - Multi Queue} and~\ref{fig:latency-bbr2-dctcp}) flows. Moreover, Prague flows do not experience extremely high latency when passing through a single queue ECN-enabled bottleneck (Figure~\ref{fig: Prague Latency vs BBRv2 - Single Queue}), unlike what is observed in the Prague vs. CUBIC coexistence case (Figure~\ref{fig: Prague Latency vs Cubic - Single Queue}).

\textit{Circumstances where Prague is harmed due to coexistence with a BBR flow.} In single queue bottlenecks with shallow buffers, Prague's throughput is drastically reduced compared to BBRv1 (Figure~\ref{fig:tput-bbr1}), as BBRv1 is not sensitive to packet loss, unlike Prague. This aligns with the well-known behavior of BBRv1 when compared to classic loss-based CCAs~\cite{bbr-2019-ware,pam-bbr3,bbr2-fluid-model-imc-22}. BBRv2, in contrast, shows better fairness in shallow buffers, due to its loss sensitivity, as documented in~\cite{pam-bbr3,bbr2-fluid-model-imc-22}. In the FIFO with ECN but without AQM, as also seen in Prague vs CUBIC, the non-ECN flow dominates the ECN flow (Figure~\ref{fig:Prague Throughput vs BBRv2 - Single Queue} and~\ref{fig:tput-bbr1}), consistent with our previous findings in~\cite{to-switch-or-not-to-switch-prague}. For BBRv1, since it neither responds to ECN nor is sensitive to losses, it causes Prague's throughput to drastically decrease when no AQM is present in an ECN-supported bottleneck (Figure~\ref{fig:tput-bbr1}). In FQ without AQM, in shallow buffers, the bottleneck fails to provide fairness and BBRv1 is harmful to Prague (Figure~\ref{fig:tput-bbr1}). In drop-based AQMs, such as PIE and CoDel, BBRv1 consistently achieves significantly higher throughput compared to Prague across all buffer sizes (Figure~\ref{fig:tput-bbr1}). This is due to Prague's loss sensitivity, in contrast to BBRv1. Even with larger buffer sizes, BBRv1 dominates because these AQMs use early-drop mechanisms based on target queueing latency, rather than tail-drop. In the case of BBRv2, for larger buffer sizes with drop-based AQMs, Prague is harmed (Figure~\ref{fig:Prague Throughput vs BBRv2 - Single Queue}, and~\ref{fig:tput-bbr2-dctcp}). This is because these AQMs use early-drop mechanisms, which increase the number of drops. Prague, effectively Reno in non-ECN bottlenecks, is more sensitive to drop events as a loss based CCA. This behaviour is observed also in~\cite{bbr2-fluid-model-imc-22}. In DualPI2, both two-sided DCTCP-style ECN BBRv2 (Figure~\ref{fig:tput-bbr2-dctcp}), and BBRv1 flows (Figure~\ref{fig:tput-bbr1}), dominate Prague. This could be due to the inherent design of this AQM type, indicating that the coexistence of BBR flows in DualPI2 AQM requires further investigation. With this bottleneck, as also seen in Prague vs CUBIC case, the ECN fallback algorithm does not perform well (Figure~\ref{fig:Prague Throughput vs BBRv2 - Multi Queue}).

\textit{Circumstances where Prague harms a BBR flow.} In single queue bottlenecks with ECN, Prague operates as a scalable CCA and generally degrades BBRv2's performance especially for large buffer sizes (Figure~\ref{fig:Prague Throughput vs BBRv2 - Single Queue}). In AQMs with ECN, the Prague flow gains a significant throughput advantage compared to the BBRv2 flow, as also seen in the Prague vs. CUBIC scenario (Figure~\ref{fig:Prague Throughput vs Cubic - Single Queue}). This is due to the more precise congestion feedback provided by AccECN and the finer adjustments made by TCP Prague compared to BBRv2. (In this case, Prague’s fallback algorithm appears to mitigate the harm.) BBRv2 with DCTCP-style marking on the receiver is similar (Figure~\ref{fig:tput-bbr2-dctcp}) but the harm is less severe due to the more refined CWND adjustments of the BBRv2 flow.

\begin{tcolorbox}[breakable, enhanced, beforeafter skip=0.5\baselineskip, before upper={\parindent15pt},colback=black!3!white,colframe=black,title=RQ1: \RQPrague]

A Prague flow is likely to share a bottleneck with CUBIC or BBR.

In bottleneck queues that drop (rather than mark) packets, or those that enforce fairness, Prague and CUBIC appear to coexist without harm.

However, single queue ECN bottlenecks are more problematic. In particular, in a FIFO queue with ECN, either the Prague flow is harmed (when the CUBIC flow does not respond to ECN) or the CUBIC flow is harmed (if it does respond to ECN). The ECN fallback heuristic in Prague mitigates the problem in the single queue bottleneck, but has new problems in the L4S DualPI2 AQM setting. 

The ultra-low latency benefits of TCP Prague are only realized with L4S-aware AQMs like FQ-CoDel and DualPI2.

In many single-queue bottleneck types, Prague does not coexist well with BBRv1/BBRv2, with either the Prague or the BBR flow capturing most of the link capacity. (This is partially, but not fully, mitigated if the BBRv2 flow uses DCTCP-style ECN marking with receiver support.)

Queues that enforce fairness are generally effective at preventing harm between Prague and BBR flows, but the DualPI2 AQM is not well tuned for BBR flows, and the Prague flow is disadvantaged as a result.
\end{tcolorbox}

\begin{figure*}
  \centering
  \includegraphics[width=0.9\textwidth]{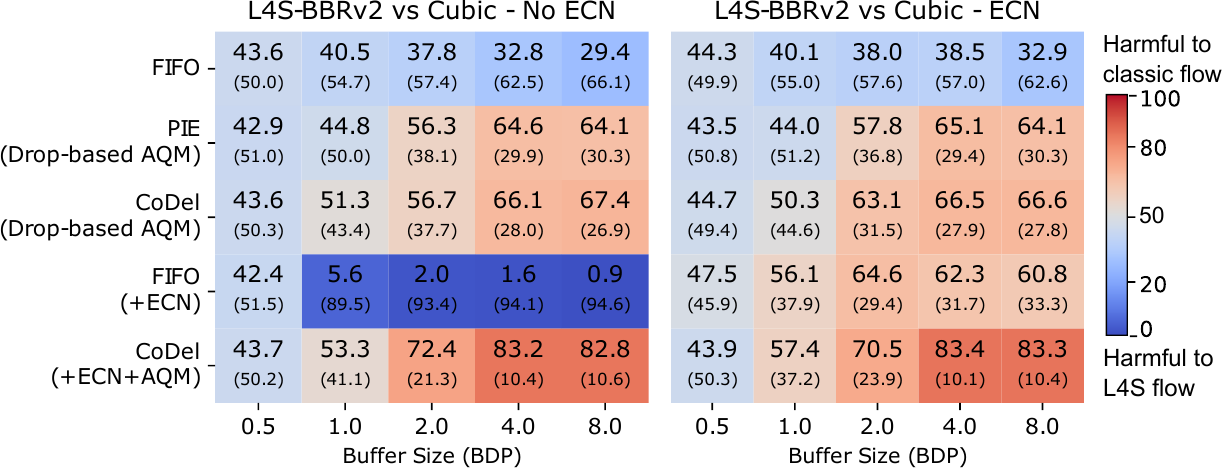}
  \caption{Throughput of L4S-compatible BBRv2 (and CUBIC, in parentheses) in Mbps when sharing a bottleneck. (single queue AQMs)}
  \label{fig:Scalable BBRv2 Throughput vs Cubic - Single Queue}
  \vspace{0.1cm}

  \centering
  \includegraphics[width=0.9\textwidth]{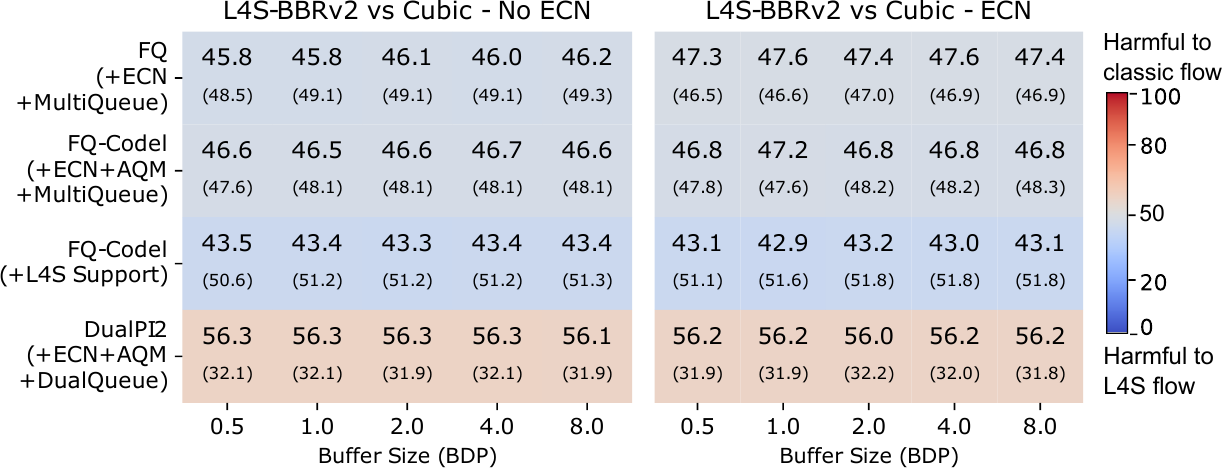}
  \caption{Throughput of L4S-compatible BBRv2 (and CUBIC, in parentheses) in Mbps when sharing a bottleneck. (multi/dual queue AQMs)}
  \label{fig:Scalable BBRv2 Throughput vs Cubic - Multi Queue}
  \vspace{0.1cm}

  \centering
  \includegraphics[width=0.9\textwidth]{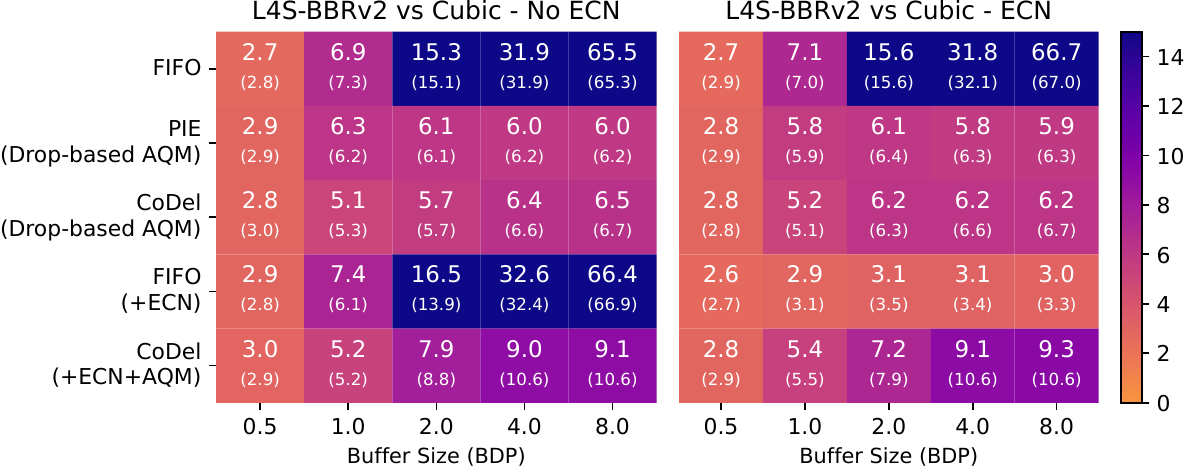}
  \caption{L4S-compatible BBRv2 (and CUBIC, in parentheses) queuing delay (ms) when sharing a bottleneck. (single queue AQMs)}
  \label{fig: Scalable BBRv2 Latency vs Cubic - Single Queue}
  \vspace{0.1cm}

  \centering
  \includegraphics[width=0.9\textwidth]{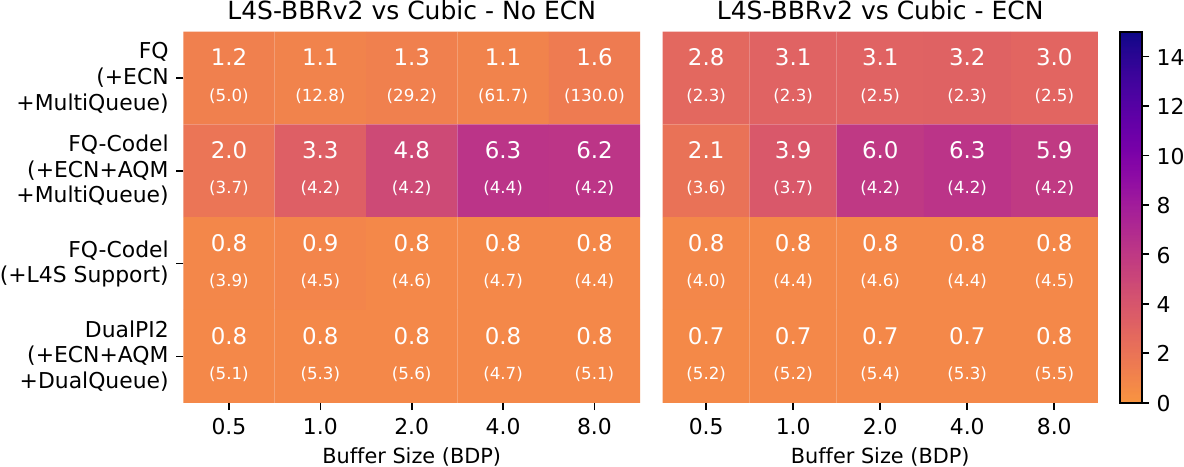}
  \caption{L4S-compatible BBRv2 (and CUBIC, in parentheses) queuing delay (ms) when sharing a bottleneck. (multi/dual queue AQMs)}
  \label{fig: Scalable BBRv2 Latency vs Cubic - Multi Queue}

\end{figure*}

\begin{figure*}

  \centering
  \includegraphics[width=1.0\textwidth]{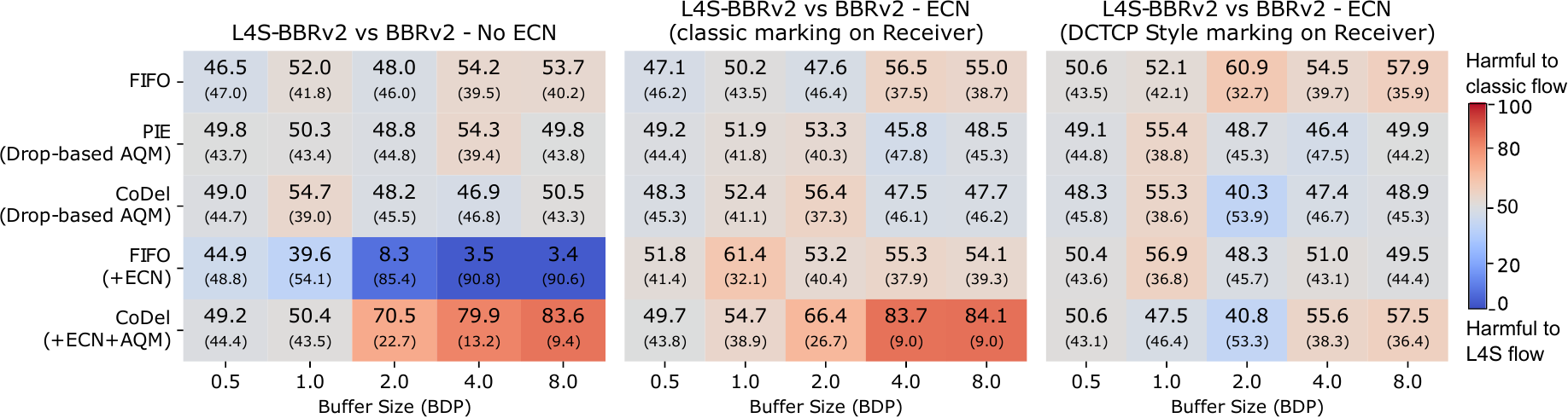}
  \caption{Throughput of L4S-compatible BBRv2 (and non-L4S-compatible BBRv2, in parentheses) in Mbps when sharing a bottleneck (single queue AQMs)}
  \label{fig:Scalable BBRv2 Throughput vs BBRv2 - Single Queue}
  \vspace{0.1cm}

  \centering
  \includegraphics[width=1.0\textwidth]{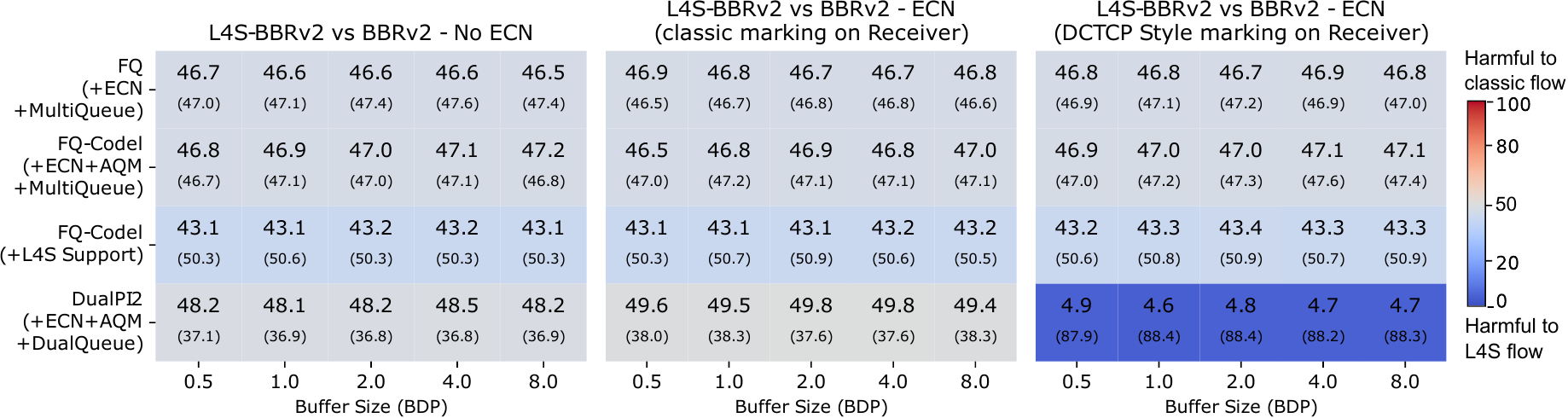}
  \caption{Throughput of L4S-compatible BBRv2 (and non-L4S-compatible BBRv2, in parentheses) in Mbps when sharing a bottleneck. (multi/dual queue AQMs)}
  \label{fig:Scalable BBRv2 Throughput vs BBRv2 - Multi Queue}
  \vspace{0.1cm}

  \centering
  \includegraphics[width=1.0\textwidth]{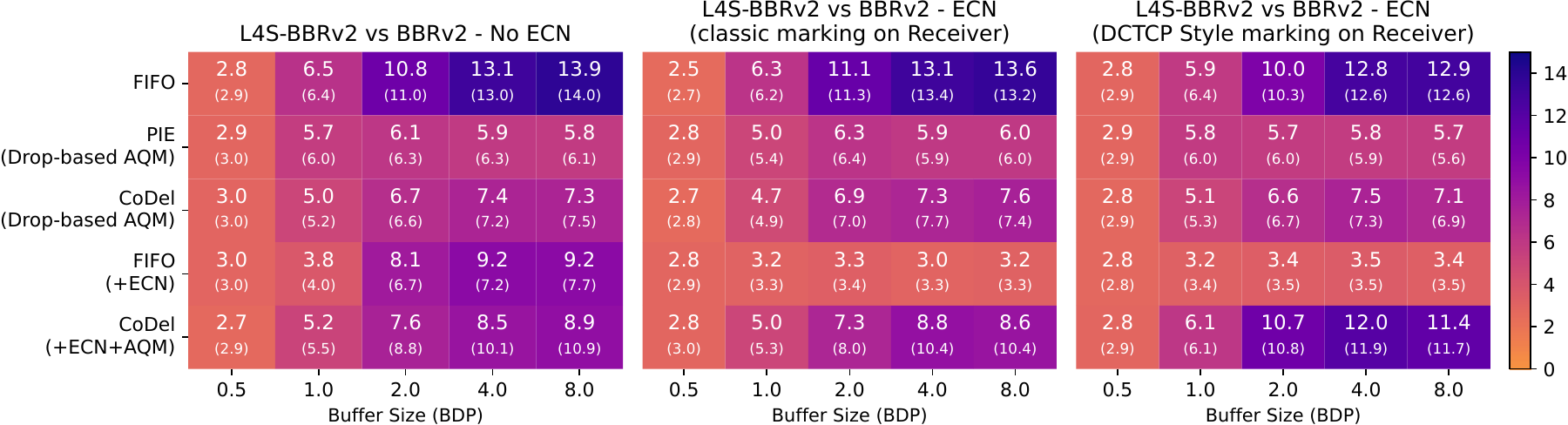}
  \caption{L4S-compatible BBRv2 (and non-L4S-compatible BBRv2, in parentheses) queuing delay (ms) when sharing a bottleneck. (single queue AQMs)}
  \label{fig: Scalable BBRv2 Latency vs BBRv2 - Single Queue}

  \centering
  \includegraphics[width=1.0\textwidth]{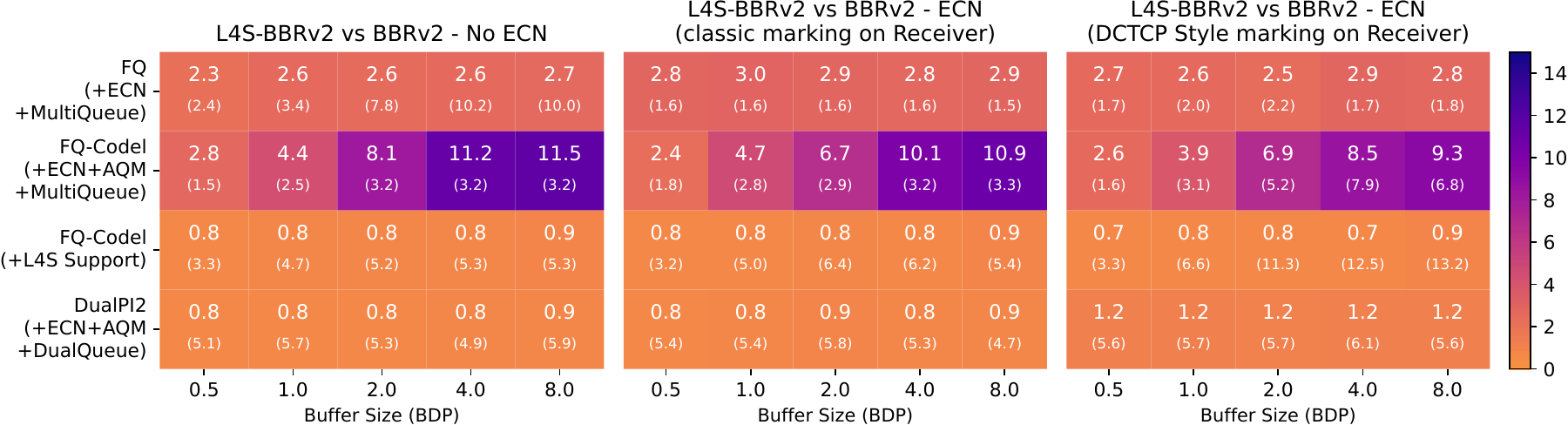}
  \caption{L4S-compatible BBRv2 (and non-L4S-compatible BBRv2, in parentheses) queuing delay (ms) when sharing bottleneck with BBRv2 flow. (multi/dual queue AQMs)}
  \label{fig: Scalable BBRv2 Latency vs BBRv2 - Multi Queue}

\end{figure*}

\vspace{-1em}

\subsection{RQ2: \RQBBR}
\raggedbottom

In this section, we evaluate the throughput and latency of an L4S-compatible BBRv2 with AccECN when it shares a bottleneck queue with a CUBIC flow, a one-sided DCTCP-style ECN (only on the sender) BBRv2 flow, and a two-sided DCTCP-style ECN BBRv2 flow. We compare this to Prague coexistence with the same types of flows, to understand whether L4S-compatible BBRv2 has more favorable properties for adoption than TCP Prague. 

The throughput of the L4S-compatible BBRv2 flow vs. the CUBIC flow (shown in parentheses) is in Figure~\ref{fig:Scalable BBRv2 Throughput vs Cubic - Single Queue}, and vs. the non-L4S-BBRv2 flow in Figure~\ref{fig:Scalable BBRv2 Throughput vs BBRv2 - Single Queue} for single queue bottleneck scenarios. For multi-queue bottleneck scenarios, the throughput is shown in Figure~\ref{fig:Scalable BBRv2 Throughput vs Cubic - Multi Queue} and Figure~\ref{fig:Scalable BBRv2 Throughput vs BBRv2 - Multi Queue}. The latency for each experiment is shown in Figure~\ref{fig: Scalable BBRv2 Latency vs Cubic - Single Queue} and Figure~\ref{fig: Scalable BBRv2 Latency vs Cubic - Multi Queue} (vs. CUBIC, in parentheses), and in Figure~\ref{fig: Scalable BBRv2 Latency vs BBRv2 - Single Queue} and Figure~\ref{fig: Scalable BBRv2 Latency vs BBRv2 - Multi Queue} (vs. non-L4S BBRv2, in parentheses).

 \textit{Circumstances where L4S-compatible BBRv2 has more favorable properties for adoption than Prague.}  L4S-compatible BBRv2 is less harmful than Prague when sharing a single ECN queue, especially a FIFO+ECN queue, with a CUBIC or BBRv2-DCTCP-style flow that is responsive to ECN (Figure~\ref{fig:Scalable BBRv2 Throughput vs Cubic - Single Queue}, and~\ref{fig:Scalable BBRv2 Throughput vs BBRv2 - Single Queue}). In drop-based bottlenecks, such as FIFO, PIE, and CoDel, L4S-compatible BBRv2 provides stable performance and neither harms nor is harmed when sharing the bottleneck with another BBRv2 flow (Figure~\ref{fig:Scalable BBRv2 Throughput vs BBRv2 - Single Queue}). However, for Prague, sharing the bottleneck with BBRv2, particularly in single-queue bottlenecks, poses challenges, as explained in the previous section. When a DualPI2 AQM bottleneck is shared with a BBRv2 flow without ECN, Prague is slightly harmed (Figure~\ref{fig:Prague Throughput vs BBRv2 - Multi Queue}), while L4S-compatible BBRv2 achieves approximately a fair share (Figure~\ref{fig:Scalable BBRv2 Throughput vs BBRv2 - Multi Queue}).

 \textit{Circumstances where L4S-compatible BBRv2 is similar to Prague.} 
In most cases, the performance differences are minimal. For example, in drop-based AQMs, both L4S flows exhibit either a slight advantage or disadvantage compared to the CUBIC flow in the bottleneck (Figure~\ref{fig:Scalable BBRv2 Throughput vs Cubic - Single Queue}). In FQ bottlenecks, both achieve roughly their fair share of throughput, and in L4S-aware bottlenecks, both maintain ultra-low latency (Figure~\ref{fig:Scalable BBRv2 Throughput vs Cubic - Multi Queue},~\ref{fig: Scalable BBRv2 Latency vs BBRv2 - Multi Queue}). When the classic queue contains BBRv1 or two-sided DCTCP-style ECN-marking BBRv2, these flows dominate both L4S-compatible BBRv2 and Prague (Figure~\ref{fig:Scalable BBRv2 Throughput vs BBRv2 - Multi Queue}, and Figure~\ref{fig:tput-bbr2-dctcp}). Although the performance of BBRv1 is not shown here, it is observed that when sharing a bottleneck with a BBRv1 flow, the performance of both L4S-compatible BBRv2 and Prague is similar across all bottleneck types (Figure~\ref{fig:tput-bbr1}). In terms of latency, both Prague and L4S-compatible BBRv2 can experience extreme queuing latencies, even with ECN enabled, when competing with CUBIC on a FIFO+ECN bottleneck (Figure~\ref{fig: Scalable BBRv2 Latency vs Cubic - Single Queue}), and both perform similarly in terms of latency when sharing the single queue bottleneck with BBRv2 (Figure~\ref{fig: Scalable BBRv2 Latency vs BBRv2 - Single Queue}).

 \textit{Circumstances where L4S-compatible BBRv2 has less favorable properties for adoption than Prague.} In DualPI2, L4S-compatible BBRv2 experiences slightly higher queuing latency compared to Prague, though both maintain queuing latency under 1ms (Figure~\ref{fig: Scalable BBRv2 Latency vs Cubic - Multi Queue}, ~\ref{fig: Scalable BBRv2 Latency vs BBRv2 - Multi Queue}). 

\begin{tcolorbox}[beforeafter skip=0.5\baselineskip, before upper={\parindent15pt},colback=black!3!white,colframe=black,title=RQ2: \RQBBR]

L4S-compatible BBRv2 is similar to Prague with respect to sharing single queue ECN-enabled bottlenecks with a CUBIC flow - neither is safe from harming, or being harmed, especially when the CUBIC flow ignores ECN. 
However, when sharing with a CUBIC flow that \emph{does} respond to ECN, L4S-compatible BBRv2 is less harmful than Prague.

When sharing a bottleneck with a non-L4S-compatible BBRv2 flow, L4S-compatible BBRv2 is generally less likely to harm or be harmed than Prague, especially in single queue and DualPI2 queues.

\end{tcolorbox}

\subsection{RQ3:  \RQMulti}
\raggedbottom

\begin{figure*}

  \centering
  \includegraphics[width=1.0\textwidth]{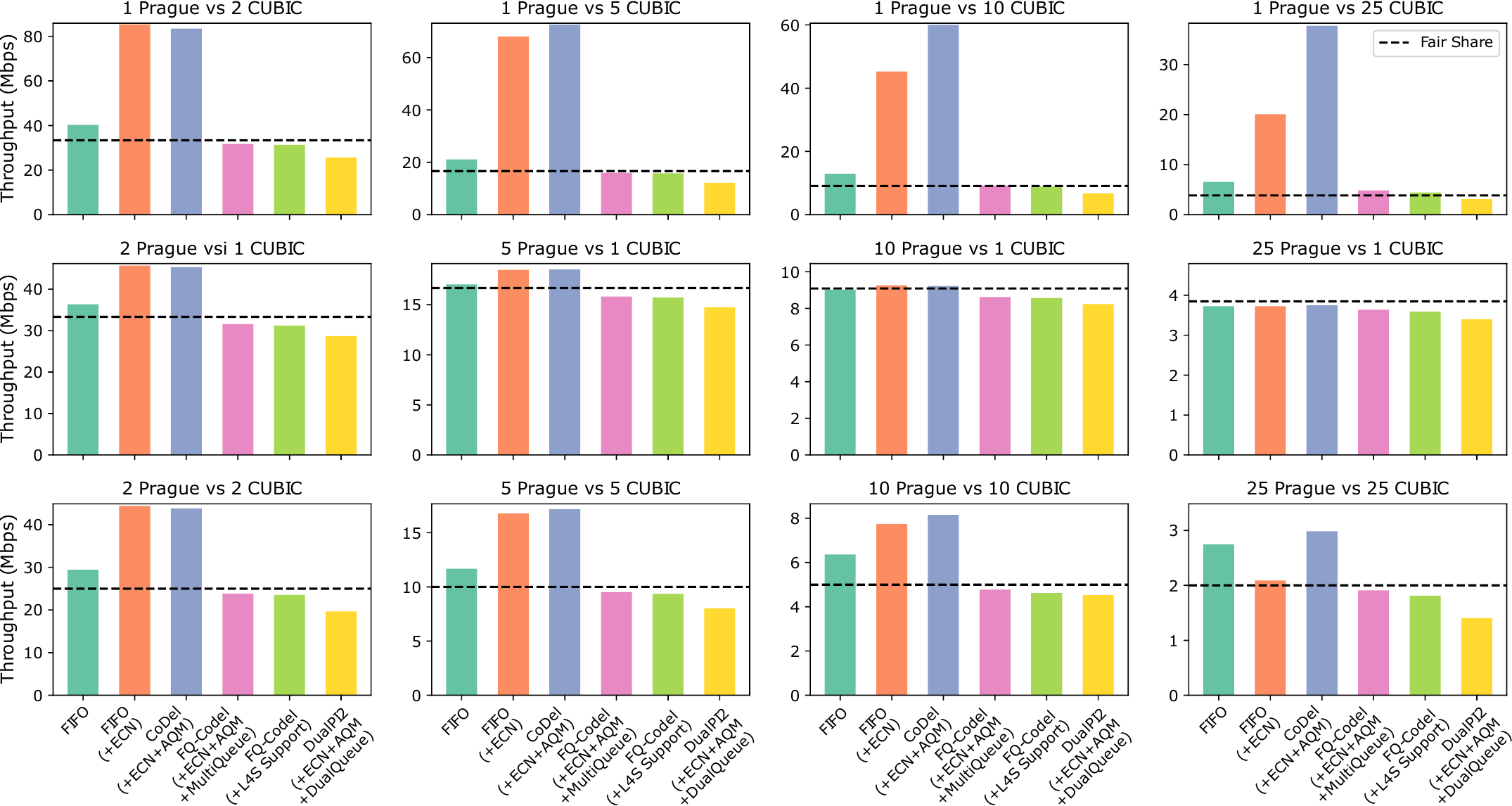}
  \caption{1 vs. many, many vs. many flow scenarios - Prague throughput (Mbps).}
  \label{fig:Prague Throughput vs multiple cubic flows}
  \vspace{0.1cm}

  \centering
  \includegraphics[width=1.0\textwidth]{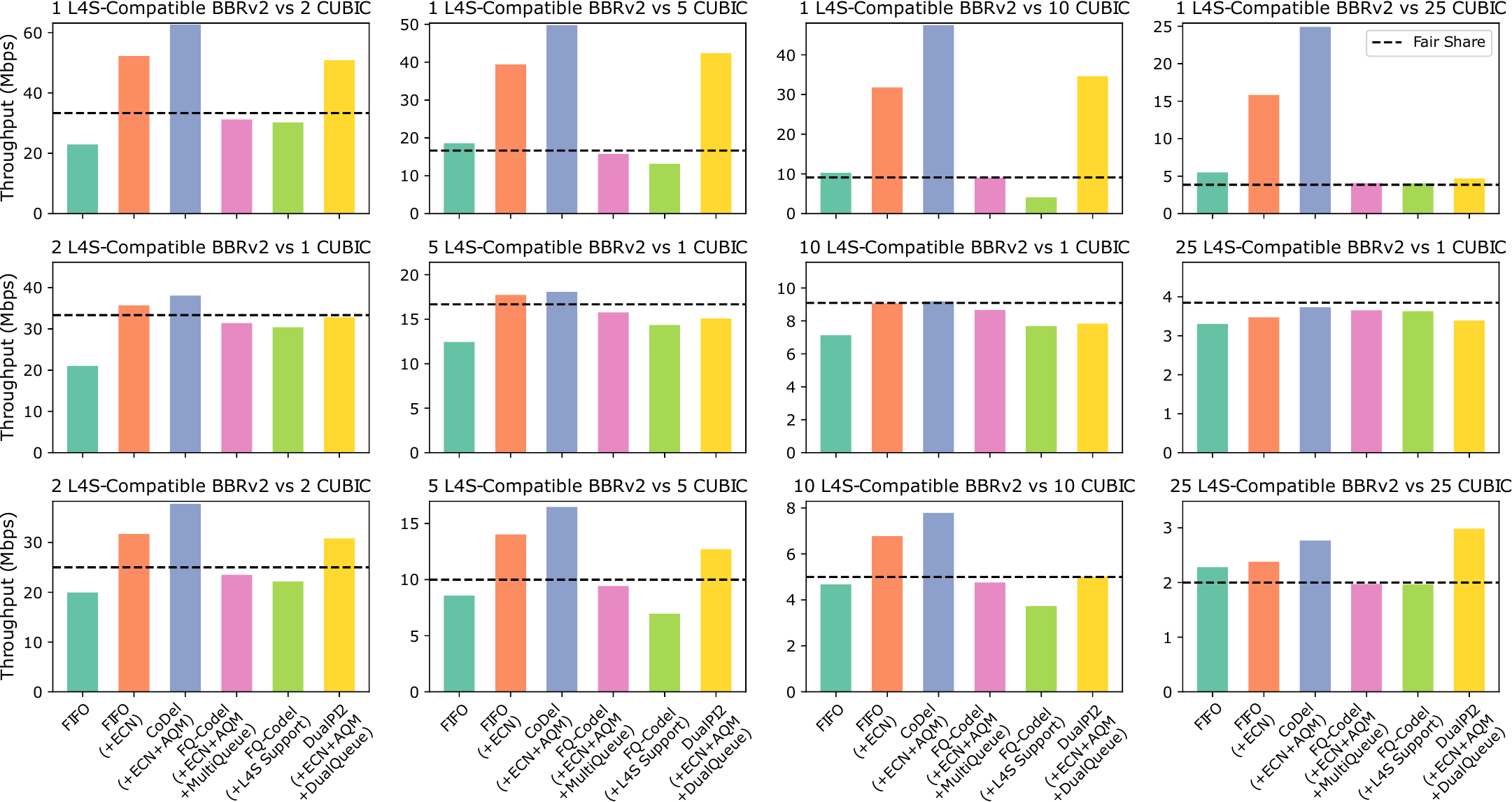}
  \caption{1 vs. many, many vs many flow scenarios - L4S-compatible BBRv2 throughput (Mbps).}
  \label{fig:Scalable BBRv2 Throughput vs multiple cubic flows}
  \vspace{0.1cm}
  
\end{figure*}

To address our third research question, we evaluate the performance of TCP Prague and L4S-compatible BBRv2 to understand the impact of the number of flows on the results previously discussed in RQ1 and RQ2 for the 1vs1 experiments. The average throughput of Prague flows  is shown in Figure~\ref{fig:Prague Throughput vs multiple cubic flows} and L4S-compatible BBRv2 flows in Figure~\ref{fig:Scalable BBRv2 Throughput vs multiple cubic flows}. 

\textit{Circumstances where the harm is not mitigated.} When a single Prague or L4S-compatible BBRv2 flow competes against multiple CUBIC flows, the L4S flow consistently achieves much higher throughput than its fair share - in the case of a single-queue ECN bottleneck, even against 25 CUBIC flows  (Figure~\ref{fig:Prague Throughput vs multiple cubic flows},~\ref{fig:Scalable BBRv2 Throughput vs multiple cubic flows}). 

In DualPI2, a single BBRv2 flow maintains a significant throughput advantage over multiple CUBIC flows, even with up to 10 CUBIC flows (Figure~\ref{fig:Scalable BBRv2 Throughput vs multiple cubic flows}). However, when competing against 25 CUBIC flows, BBRv2 achieves a fair share of the throughput. This differs from Prague, where Prague flows generally achieve close to their fair share or slightly less across all settings with DualPI2.

\textit{Circumstances where the harm is mitigated.} Across all AQM types, the Prague flows tend to achieve their fair share in the 25 Prague vs. 25 CUBIC flows scenario, as illustrated in Figure~\ref{fig:Prague Throughput vs multiple cubic flows}, showing they are not harmed by the CUBIC flows. On the other hand, with a high number of flows, such as 25 BBRv2 vs. 25 CUBIC, particularly with L4S-aware FQ-CoDel, BBRv2 flows experience around 10 ms of queuing latency, although not shown in the figure, which is significantly higher than their 1 ms target delay. This behavior could vary under different network configurations. However, this represents a possible scenario where the ultra-low latency benefit is not realized, demonstrating that even with an L4S-aware FQ-CoDel bottleneck, L4S flows can experience high queuing latency.

\begin{tcolorbox}[beforeafter skip=0.5\baselineskip, before upper={\parindent15pt},colback=black!3!white,colframe=black,title=RQ3: \RQMulti]

When \emph{one} Prague or L4S-compatible BBRv2 flow shares a single-queue ECN bottleneck with \emph{many} CUBIC flows, it captures $5-10\times$ its fair share of the link capacity, harming the CUBIC flows. 

However, when multiple Prague flows share a bottleneck with one CUBIC flow, or when multiple Prague and multiple CUBIC flows share a bottleneck, some of the issues described in Section~\ref{sec:RQ1} are mitigated.

For BBRv2, aside from the issue noted above, as the number of flows increases—whether 1 BBRv2 vs. 25 CUBIC, 25 CUBIC vs. 1 BBRv2, or more flows—the harmful effects are mostly mitigated.

When multiple L4S-compatible BBRv2 flows share an L4S-aware FQ bottleneck with multiple CUBIC flows, it is not guaranteed that the L4S flows will achieve ultra-low latency especially with high number of flows.

\end{tcolorbox}

\begin{figure*}

  \centering
  \includegraphics[width=0.9\textwidth]{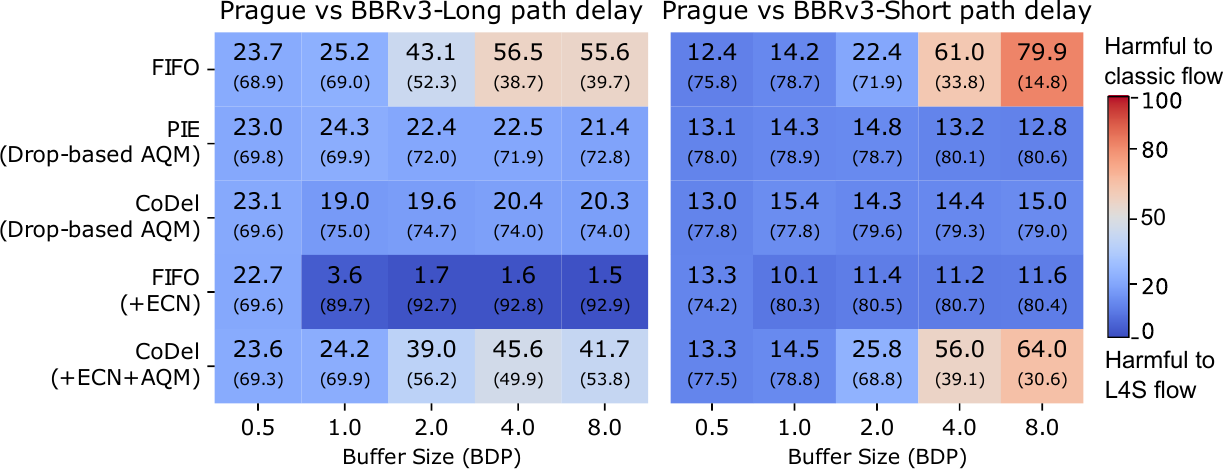}
  \caption{Throughput of Prague (and BBRv3, in parentheses) in Mbps when sharing a bottleneck (single queue AQMs) 
  }
  \label{fig:Prague Throughput vs BBRv3 - Single Queue}
  \vspace{0.1cm}

  \centering
  \includegraphics[width=0.9\textwidth]{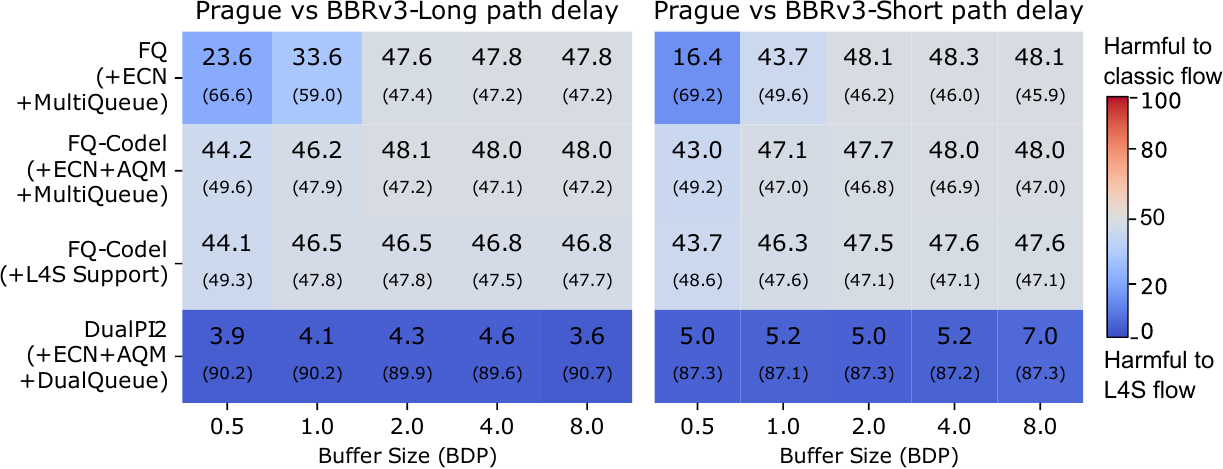}
  \caption{Throughput of Prague (and BBRv3, in parentheses) in Mbps when sharing a bottleneck(multi/dual queue AQMs) 
  }
  \label{fig:Prague Throughput vs BBRv3 - Multi Queue}
  \vspace{0.1cm}

  \centering
  \includegraphics[width=0.9\textwidth]{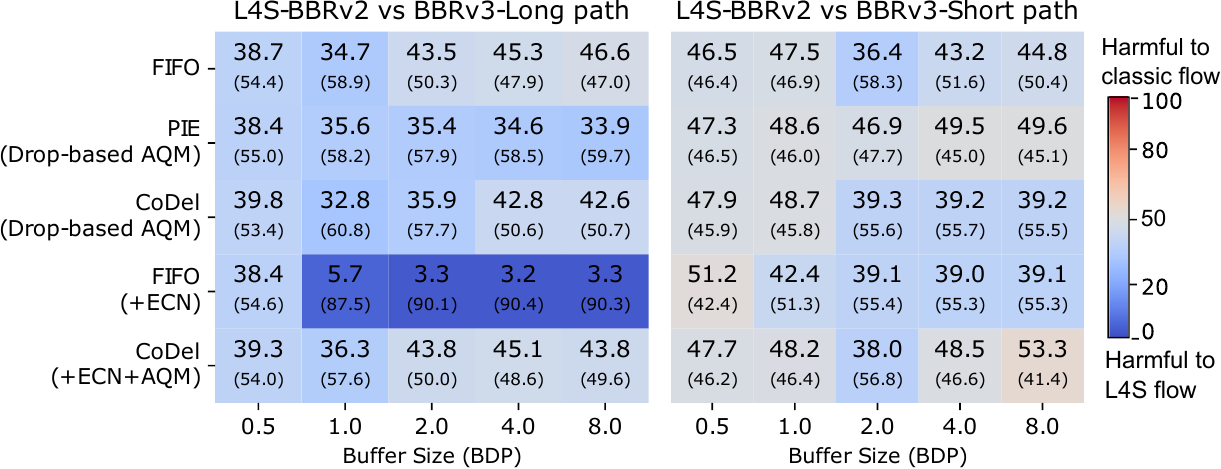}
  \caption{Throughput of L4S-compatible BBRv2 (and BBRv3, in parentheses) in Mbps when sharing a bottleneck (single queue AQMs) 
  }
  \label{fig:l4s-bbrv2 Throughput vs BBRv3 - Single Queue}
  \vspace{0.1cm}

  \centering
  \includegraphics[width=0.9\textwidth]{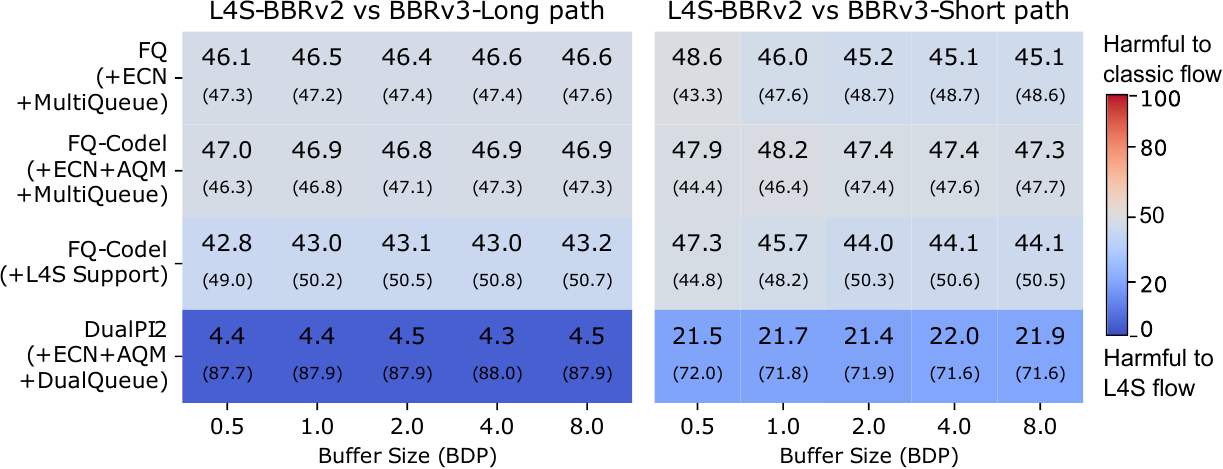}
  \caption{Throughput of L4S-compatible BBRv2 (and BBRv3, in parentheses) in Mbps when sharing a bottleneck(multi/dual queue AQMs) 
  }
  \label{fig:l4s-bbrv2 Throughput vs BBRv3 - Multi Queue}
  \vspace{0.1cm}
\end{figure*}

\subsection{RQ4: \RQBBRThree} ~\label{subsec:EvalBBRv3}
In this section, we evaluate the coexistence of BBRv3 with Prague and L4S-compatible BBRv2. BBRv3's ECN behavior depends on the measured minimum RTT on the path. Even if the sender and receiver enable ECN, if the minRTT exceeds 5ms, the BBR sender does not respond to ECN. We evaluate this with a 10ms RTT, where ECN is not enabled effectively, and a 2~ms RTT, where ECN is enabled and the threshold is 1ms. We use a one-sided DCTCP-style ECN mechanism, with the receiver using classic ECN marking. Since the sender does not know the ECN marking types it will encounter, we consider arbitrary endpoints without specialized support, as would be the case in a real network scenario.
The throughput of the Prague flow (and the BBRv3 flow, in parenthesis) is shown in Figures~\ref{fig:Prague Throughput vs BBRv3 - Single Queue} (single-queue), and Figure~\ref{fig:Prague Throughput vs BBRv3 - Multi Queue} (multi-queue). Similarly, the throughput of the L4S-compatible BBRv2 flow (and the BBRv3 flow, in parenthesis) is shown in Figures~\ref{fig:l4s-bbrv2 Throughput vs BBRv3 - Single Queue} (single-queue), and Figure~\ref{fig:l4s-bbrv2 Throughput vs BBRv3 - Multi Queue} (multi-queue).

\textit{Circumstances where the BBRv3 approach more favorable.} In FIFO+ECN, the harm caused by BBRv3 to Prague is less severe with short path delays (with ECN) compared to long path delays (without ECN) (Figure~\ref{fig:Prague Throughput vs BBRv3 - Single Queue}). Beyond this, there is no significant performance advantage for BBRv3's approach in the coexistence of Prague and BBRv3. In multiple-queue bottlenecks, the performance is very similar in both cases (Figure~\ref{fig:Prague Throughput vs BBRv3 - Multi Queue}).

On the other hand, with L4S-compatible BBRv2 in short path delays, the harmful effects in FIFO+ECN are completely mitigated (Figure~\ref{fig:l4s-bbrv2 Throughput vs BBRv3 - Single Queue}). Furthermore, in DualPI2, the harm to BBR flows is less severe in short path delays with ECN compared to long path delays without ECN (Figure~\ref{fig:l4s-bbrv2 Throughput vs BBRv3 - Multi Queue}).

\textit{Circumstances where the BBRv3 approach is not favorable.} There is no significant indication that this approach is less favorable.

\begin{tcolorbox}[beforeafter skip=0.5\baselineskip,  before upper={\parindent15pt},colback=black!3!white,colframe=black,title=RQ4: \RQBBRThree]

For several types of single queue bottlenecks, and DualPI2, a Prague flow is harmed due to coexistence with a BBRv3 flow, whether the BBRv3 flow is responsive to ECN (short path delay) or not (long path delay).

L4S-compatible BBRv2 generally coexists well with BBRv3 in either short or long path delay cases. The exceptions are FIFO+ECN with long path delays (no ECN), where BBRv3 harms the L4S-compatible BBRv2 flow — an issue mitigated with short path delays — and DualPI2 for both paths, where BBRv3 again harms the L4S-compatible BBRv2 flow.

\end{tcolorbox}
\vspace{-1em}


\section{Conclusion}
\label{sec:conclusion}

In this work, we investigate the behavior that a sender may expect upon adopting an L4S-compatible congestion control, with a systematic evaluation in a setting based on a residential broadband network, considering a range of queue types and types of non-L4S flows at the shared bottleneck. We identify circumstances in which the L4S flow gets only a small fraction of the bottleneck link bandwidth, while the non-L4S flow dominates. We also identify circumstances in which the L4S flow is harmful to the non-L4S flow, capturing most of the bottleneck link bandwidth. Furthermore, we find that this effect is not always mitigated when more flows share the bottleneck link. Given that the sender of an L4S flow cannot be sure what type of queue is at the bottleneck or what other flows will share that queue, a range of outcomes for throughput and latency are possible.

Some key results are summarized in Table~\ref{tab:my-table}, and below:

\begin{itemize}
\item \textbf{In many settings with a single queue, the L4S flow harms or is harmed by the classic flow.} These outcomes depend on the behavior of the competing flow (e.g., whether it supports ECN or not, and whether it uses CUBIC or BBR). This effect is not necessarily mitigated when multiple flows are sharing the bottleneck link. 
\item \textbf{The DualPI2 AQM, which is designed specifically for L4S, is not well-tuned for sharing a link with BBR flows.} This finding is concerning because of the (increasing) prevalence of BBR on the Internet~\cite{census}.
\item \textbf{The low latency benefits of L4S are realized only when the bottleneck queue isolates flows - with a fair queue or dual queue AQM.} This is by design, and is not a surprising result, but in combination with the other problems, further disincentivizes the adoption of L4S.
\end{itemize}

\vspace{-2em}
\begin{table}
\centering
\caption{Can a sender turn on L4S (\cmark) or would it harm/be harmed if it did (\xmark)? (\faBolt~indicates settings where ultra-low latency is achieved, otherwise it is not.)}
\resizebox{\textwidth}{!}{%
\renewcommand{\arraystretch}{1.1} 
\begin{tabular}{|c|c|c|c|c|c|}
\hline
AQM & CUBIC & BBRv1 & \begin{tabular}[c]{@{}c@{}}BBRv2 (one-\\ \;sided DCTCP ECN)\;\end{tabular} & \begin{tabular}[c]{@{}c@{}}BBRv2 (two-\\ \;sided DCTCP ECN)\;\end{tabular} & BBRv3 \\
\hline
SQ (Drop) & \begin{tabular}[c]{@{}c@{}}\cmark \\ (Fig.~\ref{fig:Prague Throughput vs Cubic - Single Queue}, \ref{fig:Scalable BBRv2 Throughput vs Cubic - Single Queue})\end{tabular} & \begin{tabular}[c]{@{}c@{}}\xmark \\ (Fig.~\ref{fig:tput-bbr1})\end{tabular} & \multicolumn{3}{c|}{\begin{tabular}[c]{@{}c@{}}\xmark~Prague \quad \quad \cmark~L4S-BBRv2\\ (Fig.~\ref{fig:Prague Throughput vs BBRv2 - Single Queue}, \ref{fig:tput-bbr2-dctcp}, \ref{fig:Scalable BBRv2 Throughput vs BBRv2 - Single Queue},~\ref{fig:Prague Throughput vs BBRv3 - Single Queue}, \ref{fig:l4s-bbrv2 Throughput vs BBRv3 - Single Queue})\end{tabular}}  \\
\hline
SQ (ECN) & \begin{tabular}[c]{@{}c@{}}\xmark \\ (Fig.~\ref{fig:Prague Throughput vs Cubic - Single Queue}, \ref{fig:Scalable BBRv2 Throughput vs Cubic - Single Queue})\end{tabular} & \begin{tabular}[c]{@{}c@{}}\xmark \\(Fig.~\ref{fig:tput-bbr1}) \end{tabular}& \begin{tabular}[c]{@{}c@{}}\xmark\\ (Fig.~\ref{fig:Prague Throughput vs BBRv2 - Single Queue}, \ref{fig:Scalable BBRv2 Throughput vs BBRv2 - Single Queue})\end{tabular} & \begin{tabular}[c]{@{}c@{}}\cmark\\ (Fig.~\ref{fig:tput-bbr2-dctcp}, \ref{fig:Scalable BBRv2 Throughput vs BBRv2 - Single Queue} )\end{tabular} & \begin{tabular}[c]{@{}c@{}}\xmark \\ (Fig.~\ref{fig:Prague Throughput vs BBRv3 - Single Queue}, \ref{fig:l4s-bbrv2 Throughput vs BBRv3 - Single Queue})\end{tabular} \\
\hline
\begin{tabular}[c]{@{}c@{}}DualPI2 \\ (\faBolt) \end{tabular}& \begin{tabular}[c]{@{}c@{}}\cmark \\ (Fig.~\ref{fig:Prague Throughput vs Cubic - Multi Queue}, \ref{fig:Scalable BBRv2 Throughput vs Cubic - Multi Queue})\end{tabular} & \begin{tabular}[c]{@{}c@{}}\xmark \\ (Fig.~\ref{fig:tput-bbr1})\end{tabular} & \begin{tabular}[c]{@{}c@{}}\cmark \\ (Fig.~\ref{fig:Prague Throughput vs BBRv2 - Multi Queue}, \ref{fig:Scalable BBRv2 Throughput vs BBRv2 - Multi Queue})\end{tabular} & \begin{tabular}[c]{@{}c@{}} \xmark \\ (Fig.~\ref{fig:tput-bbr2-dctcp}, \ref{fig:Scalable BBRv2 Throughput vs BBRv2 - Multi Queue})\end{tabular} & \begin{tabular}[c]{@{}c@{}}\xmark\\ (Fig.~\ref{fig:Prague Throughput vs BBRv3 - Multi Queue}, \ref{fig:l4s-bbrv2 Throughput vs BBRv3 - Multi Queue})\end{tabular} \\
\hline
\begin{tabular}[c]{@{}c@{}}FQ \\ (\faBolt~if L4S-aware)\end{tabular} & \multicolumn{5}{c|}{\begin{tabular}[c]{@{}c@{}}\cmark \\ (Fig.~\ref{fig:Prague Throughput vs Cubic - Multi Queue}, \ref{fig:tput-bbr1}, \ref{fig:Prague Throughput vs BBRv2 - Multi Queue}, \ref{fig:Prague Throughput vs BBRv3 - Multi Queue}, \ref{fig:tput-bbr2-dctcp}, \ref{fig:Scalable BBRv2 Throughput vs Cubic - Multi Queue}, \ref{fig:Scalable BBRv2 Throughput vs BBRv2 - Multi Queue}, \ref{fig:l4s-bbrv2 Throughput vs BBRv3 - Multi Queue})\end{tabular}} \\
\hline
\end{tabular}%
}
\vspace{1em}
\label{tab:my-table}
\end{table}
\vspace{-2em}

This combination of factors - the possibility of being worse off with L4S than without, the possibility of causing harm to non-L4S flows, and that low-latency benefits are realized only in limited circumstances - may depress the adoption of L4S. As future work, we hope to investigate more directly the expected impact on L4S adoption, and strategies for mitigation.

In this study, we focused on a single network configuration with a base RTT of 10~ms and a bottleneck link capacity of 100~Mbps, to understand the residential broadband setting in depth. In real-world network paths, L4S flows will encounter a wide variety of network configurations, both in terms of the various ECN and AQM settings that were considered in this work, and different link capacities and delays not considered in this work. Some of these settings may be more favorable to L4S deployment than the one considered in our experiments. However, different results in other settings would not address the fundamental problem suggested by our results: \textbf{outside of a controlled environment, an L4S flow is reasonably likely to encounter a setting where it is harmed or causes harm.} Because an L4S flow cannot anticipate \emph{which} setting it will encounter, the existence of some problematic behaviors in our ``moderate'' residential broadband network setting is concerning. As future work, we intend to extend our evaluation to include a broader range of network settings and more realistic traffic patterns. Additionally, our evaluation primarily focused on throughput and latency, while other metrics, such as jitter and flow completion time, could offer further insights. Future work will also explore these aspects.

\begingroup
\small
\subsubsection*{Acknowledgements.} This research was supported by the New York State Center for Advanced Technology in Telecommunications and Distributed Systems (CATT), NYU WIRELESS, and the National Science Foundation (NSF) under Grant No. CNS-2148309 and OAC-2226408.
\subsubsection*{Disclosure of Interests.} The
authors have no competing interests to declare that are relevant to the content of this article.
\endgroup

\newpage

\appendix

\section{Ethical considerations}

This paper does not raise any ethical issues.

\section{Artifact description}\label{sec:appendix:artifact}

We have made the artifacts used to run the FABRIC experiments and generate the figures in this paper publicly available~\cite{our_artifacts_github_repo}.

To reproduce the results presented in this paper, shown in Figure~\ref{fig:Prague Throughput vs Cubic - Single Queue} to Figure~\ref{fig:l4s-bbrv2 Throughput vs BBRv3 - Multi Queue}, please use the Jupyter notebooks \texttt{single\_flow\_experiments.ipynb} and \texttt{multiple\_flow\_experiments.ipynb}. These notebooks run the FABRIC experiments for single-flow and multiple-flow scenarios and generate the resulting data in JSON format.

The generated JSON files can then be used with the data analysis Jupyter notebooks to reproduce the figures in this paper. Detailed instructions for running the experiments and generating the figures are included in the notebooks.

\bibliographystyle{IEEEtran}
\bibliography{ref.bib}

\begin{thebibliography}{10}
\providecommand{\url}[1]{#1}
\csname url@samestyle\endcsname
\providecommand{\newblock}{\relax}
\providecommand{\bibinfo}[2]{#2}
\providecommand{\BIBentrySTDinterwordspacing}{\spaceskip=0pt\relax}
\providecommand{\BIBentryALTinterwordstretchfactor}{4}
\providecommand{\BIBentryALTinterwordspacing}{\spaceskip=\fontdimen2\font plus
\BIBentryALTinterwordstretchfactor\fontdimen3\font minus \fontdimen4\font\relax}
\providecommand{\BIBforeignlanguage}[2]{{%
\expandafter\ifx\csname l@#1\endcsname\relax
\typeout{** WARNING: IEEEtran.bst: No hyphenation pattern has been}%
\typeout{** loaded for the language `#1'. Using the pattern for}%
\typeout{** the default language instead.}%
\else
\language=\csname l@#1\endcsname
\fi
#2}}
\providecommand{\BIBdecl}{\relax}
\BIBdecl

\bibitem{l4sarch-rfc9330}
\BIBentryALTinterwordspacing
B.~Briscoe, K.~D. Schepper, M.~Bagnulo, and G.~White, ``{Low Latency, Low Loss, and Scalable Throughput (L4S) Internet Service: Architecture},'' RFC 9330, Jan. 2023. [Online]. Available: \url{https://datatracker.ietf.org/doc/html/rfc9330}
\BIBentrySTDinterwordspacing

\bibitem{EricssonDeutscheTelekom2021}
P.~Willars, E.~Wittenmark, H.~Ronkainen, C.~Östberg, I.~Johansson, J.~Strand, P.~Lédl, and D.~Schnieders, ``{Enabling Time-Critical Applications Over 5G with Rate Adaptation},'' Ericsson and Deutsche Telekom, Tech. Rep. BNEW-21:025455 Uen, May 2021, white Paper.

\bibitem{comcast2023lowlatency}
\BIBentryALTinterwordspacing
Comcast. (2023, June) {Comcast Kicks Off Industry's First Low Latency DOCSIS Field Trials}. [Online]. Available: \url{https://corporate.comcast.com/stories/comcast-kicks-off-industrys-first-low-latency-docsis-field-trials}
\BIBentrySTDinterwordspacing

\bibitem{nokia-hololight}
{Nokia Corporation}, ``{Nokia collaborates with Hololight to deliver reliable immersive {XR} experiences with latency-improving technology L4S}, month = {November}, year = {2023}.''

\bibitem{ietf-l4s-deployment-comcast-ietf-121}
J.~Livingood, ``{Dual Queue Low Latency Networking Update},'' \url{https://datatracker.ietf.org/meeting/121/materials/slides-121-tsvwg-sessb-41-deployment-experience-01.pdf}, Nov. 2024.

\bibitem{IETF-120-understanding-prague}
\BIBentryALTinterwordspacing
K.~D. Schepper, ``{Understanding Prague for L4S},'' Presented at IETF 120, 2024. [Online]. Available: \url{https://datatracker.ietf.org/meeting/120/materials/slides-120-iccrg-understanding-prague-for-l4s-00}
\BIBentrySTDinterwordspacing

\bibitem{livingood-low-latency-deployment-07}
\BIBentryALTinterwordspacing
J.~Livingood, ``{ISP Dual Queue Networking Deployment Recommendations},'' Internet Engineering Task Force, Internet-Draft draft-livingood-low-latency-deployment-07, Oct. 2024, {Work in Progress}. [Online]. Available: \url{https://datatracker.ietf.org/doc/draft-livingood-low-latency-deployment/07/}
\BIBentrySTDinterwordspacing

\bibitem{WWDC2023-L4S-apple}
{Apple Inc.}, ``{Reduce network delays with L4S},'' {WWDC 2023, Apple Developer Videos}, 2023, available at: \url{https://developer.apple.com/videos/play/wwdc2023/10004}.

\bibitem{l4s-interop}
G.~White, ``{L4S Interop Lays Groundwork for 10G Metaverse},'' \url{https://www.cablelabs.com/blog/l4s-interop-lays-groundwork-for-10g-metaverse}, August 2022.

\bibitem{LowLatencyDOCSIS}
{CableLabs}, ``{{Low Latency DOCSIS}},'' {CableLabs Technologies}, 2024, available at: \url{https://www.cablelabs.com/technologies/low-latency-docsis}.

\bibitem{nvidia}
N.~Corporation, ``{What is the {L4S} setting in the GeForce NOW streaming quality menu?}'' \url{https://nvidia.custhelp.com/app/answers/detail/a_id/5522}.

\bibitem{prague-draft-rfc9330}
\BIBentryALTinterwordspacing
K.~D. Schepper, O.~Tilmans, B.~Briscoe, and V.~Goel, ``{Prague Congestion Control},'' Internet Engineering Task Force, Internet-Draft draft-briscoe-iccrg-prague-congestion-control-03, Oct. 2023, work in Progress. [Online]. Available: \url{https://datatracker.ietf.org/doc/draft-briscoe-iccrg-prague-congestion-control/03/}
\BIBentrySTDinterwordspacing

\bibitem{prague-paper}
B.~Briscoe, K.~De~Schepper, O.~Tilmans, M.~K{\"u}hlewind, J.~Misund, O.~Albisser, and A.~S. Ahmed, ``{Implementing the ``Prague Requirements'' for Low Latency Low Loss Scalable Throughput (L4S)},'' \emph{Netdev 0x13}, 2019.

\bibitem{accecn-draft}
\BIBentryALTinterwordspacing
B.~Briscoe, M.~Kühlewind, and R.~Scheffenegger, ``{More Accurate Explicit Congestion Notification (ECN) Feedback in TCP},'' Internet Engineering Task Force, Internet-Draft draft-ietf-tcpm-accurate-ecn-28, Nov. 2023, work in Progress. [Online]. Available: \url{https://datatracker.ietf.org/doc/draft-ietf-tcpm-accurate-ecn/28/}
\BIBentrySTDinterwordspacing

\bibitem{accecn-rfc7560}
\BIBentryALTinterwordspacing
M.~Kühlewind, R.~Scheffenegger, and B.~Briscoe, ``{Problem Statement and Requirements for Increased Accuracy in Explicit Congestion Notification (ECN) Feedback},'' RFC 7560, Aug. 2015. [Online]. Available: \url{https://datatracker.ietf.org/doc/html/rfc7560}
\BIBentrySTDinterwordspacing

\bibitem{dualpi-paper}
O.~Albisser, K.~De~Schepper, B.~Briscoe, O.~Tilmans, and H.~Steen, ``{DUALPI2—Low Latency, Low Loss and Scalable (L4S) AQM},'' NetDev 0x13, Prague, 2019.

\bibitem{dualpi-paper2}
\BIBentryALTinterwordspacing
K.~D. Schepper, O.~Albisser, O.~Tilmans, and B.~Briscoe, ``{Dual Queue Coupled AQM: Deployable Very Low Queuing Delay for All},'' 2022. [Online]. Available: \url{https://arxiv.org/abs/2209.01078}
\BIBentrySTDinterwordspacing

\bibitem{dualpi-rfc9332}
\BIBentryALTinterwordspacing
K.~D. Schepper, B.~Briscoe, and G.~White, ``{Dual-Queue Coupled Active Queue Management (AQM) for Low Latency, Low Loss, and Scalable Throughput (L4S)},'' RFC 9332, Jan. 2023. [Online]. Available: \url{https://datatracker.ietf.org/doc/html/rfc9332}
\BIBentrySTDinterwordspacing

\bibitem{to-switch-or-not-to-switch-prague}
\BIBentryALTinterwordspacing
F.~B. Sarpkaya, A.~Srivastava, F.~Fund, and S.~Panwar, ``{To switch or not to switch to TCP Prague? Incentives for adoption in a partial L4S deployment},'' in \emph{Proceedings of the 2024 Applied Networking Research Workshop}, ser. ANRW '24.\hskip 1em plus 0.5em minus 0.4em\relax New York, NY, USA: Association for Computing Machinery, 2024, p. 45–52. [Online]. Available: \url{https://doi.org/10.1145/3673422.3674896}
\BIBentrySTDinterwordspacing

\bibitem{FABRIC}
I.~Baldin, A.~Nikolich, J.~Griffioen, I.~I.~S. Monga, K.-C. Wang, T.~Lehman, and P.~Ruth, ``{FABRIC: A National-Scale Programmable Experimental Network Infrastructure},'' \emph{IEEE Internet Computing}, vol.~23, no.~6, pp. 38--47, 2019.

\bibitem{github}
F.~F. Fatih Berkay~Sarpkaya, ``{Reproducing ``Scalable Congestion Control Resolves the Delay Utilization Dilemma''},'' \url{https://github.com/fatihsarpkaya/TCP-ECN}, 2024.

\bibitem{ecn-rfc3168}
\BIBentryALTinterwordspacing
S.~Floyd, D.~K.~K. Ramakrishnan, and D.~L. Black, ``{The Addition of Explicit Congestion Notification (ECN) to IP},'' RFC 3168, Sep. 2001. [Online]. Available: \url{https://datatracker.ietf.org/doc/html/rfc3168}
\BIBentrySTDinterwordspacing

\bibitem{dctcp}
\BIBentryALTinterwordspacing
M.~Alizadeh, A.~Greenberg, D.~A. Maltz, J.~Padhye, P.~Patel, B.~Prabhakar, S.~Sengupta, and M.~Sridharan, ``{Data Center TCP (DCTCP)},'' \emph{SIGCOMM Comput. Commun. Rev.}, vol.~40, no.~4, p. 63–74, Aug. 2010. [Online]. Available: \url{https://doi.org/10.1145/1851275.1851192}
\BIBentrySTDinterwordspacing

\bibitem{ietf-tcpm-accurate-ecn-30}
\BIBentryALTinterwordspacing
B.~Briscoe, M.~Kühlewind, and R.~Scheffenegger, ``{More Accurate Explicit Congestion Notification ({ECN}) Feedback in TCP},'' Internet Engineering Task Force, Internet-Draft draft-ietf-tcpm-accurate-ecn-30, July 2024, work in Progress. [Online]. Available: \url{https://datatracker.ietf.org/doc/draft-ietf-tcpm-accurate-ecn/30/}
\BIBentrySTDinterwordspacing

\bibitem{hoeiland2018flow}
T.~Hoeiland-Joergensen, P.~McKenney, D.~Taht, J.~Gettys, and E.~Dumazet, ``{The Flow Queue CoDel Packet Scheduler and Active Queue Management Algorithm},'' RFC 8290, 2018.

\bibitem{fq_codel_l4s}
T.~Høiland-Jørgensen, ``{FQ\_Codel: Generalise ce\_threshold marking for subset of traffic},'' \url{https://git.kernel.org/pub/scm/linux/kernel/git/netdev/net-next.git/commit/?id=dfcb63ce1de6b10b}, 2021, accessed: 2024-10-01.

\bibitem{ietf-operational-guidance}
\BIBentryALTinterwordspacing
G.~White, ``{Operational Guidance on Coexistence with Classic ECN during L4S Deployment},'' Internet Engineering Task Force, Internet-Draft draft-ietf-tsvwg-l4sops-06, Mar. 2024, work in Progress. [Online]. Available: \url{https://datatracker.ietf.org/doc/draft-ietf-tsvwg-l4sops/06/}
\BIBentrySTDinterwordspacing

\bibitem{heist2021ecn}
\BIBentryALTinterwordspacing
P.~Heist and J.~Morton, ``Explicit congestion notification ({ECN}) deployment observations,'' Work in Progress, Internet-Draft, March 2021. [Online]. Available: \url{https://www.ietf.org/archive/id/draft-heist-tsvwg-ecn-deployment-observations-02.html}
\BIBentrySTDinterwordspacing

\bibitem{lim-ecn-traversal}
H.~Lim, S.~Kim, J.~Sippe, J.~Kim, G.~White, C.-H. Lee, E.~Wustrow, K.~Lee, D.~Grunwald, and S.~Ha, ``{A Fresh Look at ECN Traversal in the Wild},'' 2022.

\bibitem{jake-ce-observations}
\BIBentryALTinterwordspacing
J.~Holland, ``{CE-marking observations},'' IETF TSVWG Mailing List, April 2020. [Online]. Available: \url{https://mailarchive.ietf.org/arch/msg/tsvwg/2tbRHphJ8K_CE6is9n7iQy-VAZM/}
\BIBentrySTDinterwordspacing

\bibitem{Chun-Xiang-ecn}
C.-X. Chen and K.~Nagaoka, ``Analysis of the state of {ECN} on the internet,'' \emph{IEICE Transactions on Information and Systems}, vol. E102-D, no.~5, pp. 910--919, 2019.

\bibitem{prague-fallback}
\BIBentryALTinterwordspacing
B.~Briscoe and A.~S. Ahmed, ``{TCP Prague Fall-back on Detection of a Classic ECN AQM},'' 2021. [Online]. Available: \url{https://arxiv.org/abs/1911.00710}
\BIBentrySTDinterwordspacing

\bibitem{l4s-tests}
P.~Heist, ``{L4S Tests},'' \url{https://github.com/heistp/l4s-tests}, 2021.

\bibitem{sce-l4s-bakeoff}
{Pete Heist}, ``{SCE-L4S Bakeoff},'' \url{https://github.com/heistp/sce-l4s-bakeoff}, 2019.

\bibitem{henderson2019l4s-l4s-testing}
\BIBentryALTinterwordspacing
T.~Henderson, O.~Tilmans, and G.~White, ``{L4S Testing},'' 2019, accessed: 2024-06-12. [Online]. Available: \url{https://l4s.cablelabs.com/l4s-testing/README.html}
\BIBentrySTDinterwordspacing

\bibitem{henderson2019l4s-issues}
\BIBentryALTinterwordspacing
{Tom Henderson, Olivier Tilmans and Greg White}, ``{Testbed and Simulation Results for TSVWG Scenarios},'' 2019, accessed: 2024-06-12. [Online]. Available: \url{https://l4s.cablelabs.com/l4s_issues.html}
\BIBentrySTDinterwordspacing

\bibitem{boruoljira2020validating}
D.~BoruOljira, K.-J. Grinnemo, A.~Brunstrom, and J.~Taheri, ``{Validating the sharing behavior and latency characteristics of the L4S architecture},'' \emph{ACM SIGCOMM Computer Communication Review}, vol.~50, no.~2, pp. 37--44, 2020.

\bibitem{jain's-fairness}
R.~Jain, D.~M. Chiu, and H.~WR, ``A quantitative measure of fairness and discrimination for resource allocation in shared computer systems,'' \emph{CoRR}, vol. cs.NI/9809099, 01 1998.

\bibitem{prudentia-mmf-fairness}
\BIBentryALTinterwordspacing
A.~A. Philip, R.~Athapathu, R.~Ware, F.~F. Mkocheko, A.~Schlomer, M.~Shou, Z.~Meng, S.~Seshan, and J.~Sherry, ``Prudentia: Findings of an internet fairness watchdog,'' in \emph{Proceedings of the ACM SIGCOMM 2024 Conference}, ser. ACM SIGCOMM '24.\hskip 1em plus 0.5em minus 0.4em\relax New York, NY, USA: Association for Computing Machinery, 2024, p. 506–520. [Online]. Available: \url{https://doi.org/10.1145/3651890.3672229}
\BIBentrySTDinterwordspacing

\bibitem{cloud-based-fairness}
\BIBentryALTinterwordspacing
X.~Xu and M.~Claypool, ``Measurement of cloud-based game streaming system response to competing {TCP} cubic or {TCP} {BBR} flows,'' in \emph{Proceedings of the 22nd ACM Internet Measurement Conference}, ser. IMC '22.\hskip 1em plus 0.5em minus 0.4em\relax New York, NY, USA: Association for Computing Machinery, 2022, p. 305–316. [Online]. Available: \url{https://doi.org/10.1145/3517745.3561464}
\BIBentrySTDinterwordspacing

\bibitem{congestion-control-wild}
I.~Kunze, J.~Rüth, and O.~Hohlfeld, ``Congestion control in the wild—investigating content provider fairness,'' \emph{IEEE Transactions on Network and Service Management}, vol.~17, no.~2, pp. 1224--1238, 2020.

\bibitem{ware-jain-fairness}
\BIBentryALTinterwordspacing
R.~Ware, M.~K. Mukerjee, S.~Seshan, and J.~Sherry, ``{Beyond Jain's Fairness Index: Setting the Bar For The Deployment of Congestion Control Algorithms},'' in \emph{{Proceedings of the 18th ACM Workshop on Hot Topics in Networks}}, ser. HotNets '19.\hskip 1em plus 0.5em minus 0.4em\relax New York, NY, USA: Association for Computing Machinery, 2019, p. 17–24. [Online]. Available: \url{https://doi.org/10.1145/3365609.3365855}
\BIBentrySTDinterwordspacing

\bibitem{bbr-vs-bbr2-harm-metric-fairness}
R.~Drucker, G.~Baraskar, A.~Balasubramanian, and A.~Gandhi, ``{BBR vs. BBRv2: A Performance Evaluation},'' in \emph{2024 16th International Conference on COMmunication Systems \& NETworkS (COMSNETS)}, 2024, pp. 379--387.

\bibitem{is-it-necessary-beyond-jain-fairness}
\BIBentryALTinterwordspacing
S.~Islam, K.~Hiorth, C.~Griwodz, and M.~Welzl, ``Is it really necessary to go beyond a fairness metric for next-generation congestion control?'' in \emph{Proceedings of the 2022 Applied Networking Research Workshop}, ser. ANRW '22.\hskip 1em plus 0.5em minus 0.4em\relax New York, NY, USA: Association for Computing Machinery, 2022. [Online]. Available: \url{https://doi.org/10.1145/3547115.3547192}
\BIBentrySTDinterwordspacing

\bibitem{rfc5033bis}
\BIBentryALTinterwordspacing
M.~Duke and G.~Fairhurst, ``{Specifying New Congestion Control Algorithms},'' Internet Engineering Task Force, Internet-Draft draft-ietf-ccwg-rfc5033bis-08, Aug. 2024, work in Progress. [Online]. Available: \url{https://datatracker.ietf.org/doc/draft-ietf-ccwg-rfc5033bis/08/}
\BIBentrySTDinterwordspacing

\bibitem{codel}
\BIBentryALTinterwordspacing
K.~Nichols and V.~Jacobson, ``{Controlling Queue Delay},'' \emph{Commun. ACM}, vol.~55, no.~7, p. 42–50, July 2012. [Online]. Available: \url{https://doi.org/10.1145/2209249.2209264}
\BIBentrySTDinterwordspacing

\bibitem{rfc8033-pie}
\BIBentryALTinterwordspacing
R.~Pan, P.~Natarajan, F.~Baker, and G.~White, ``{Proportional Integral Controller Enhanced (PIE): A Lightweight Control Scheme to Address the Bufferbloat Problem},'' RFC 8033, Feb. 2017. [Online]. Available: \url{https://www.rfc-editor.org/info/rfc8033}
\BIBentrySTDinterwordspacing

\bibitem{l4srepo}
{L4S development hub}, ``{Linux kernel tree with L4S patches},'' \url{https://github.com/L4STeam/linux}, 2024.

\bibitem{cubic}
\BIBentryALTinterwordspacing
S.~Ha, I.~Rhee, and L.~Xu, ``{CUBIC: a new TCP-friendly high-speed TCP variant},'' \emph{SIGOPS Oper. Syst. Rev.}, vol.~42, no.~5, p. 64–74, July 2008. [Online]. Available: \url{https://doi.org/10.1145/1400097.1400105}
\BIBentrySTDinterwordspacing

\bibitem{census}
\BIBentryALTinterwordspacing
A.~Mishra, X.~Sun, A.~Jain, S.~Pande, R.~Joshi, and B.~Leong, ``{The Great Internet TCP Congestion Control Census},'' \emph{Proc. ACM Meas. Anal. Comput. Syst.}, vol.~3, no.~3, December 2019. [Online]. Available: \url{https://doi.org/10.1145/3366693}
\BIBentrySTDinterwordspacing

\bibitem{cubic-prevalent-ietf-rfc-9438}
\BIBentryALTinterwordspacing
L.~Xu, S.~Ha, I.~Rhee, V.~Goel, and L.~Eggert, ``{CUBIC for Fast and Long-Distance Networks},'' RFC 9438, Aug. 2023. [Online]. Available: \url{https://www.rfc-editor.org/info/rfc9438}
\BIBentrySTDinterwordspacing

\bibitem{cardwell2016bbr}
N.~Cardwell, Y.~Cheng, C.~S. Gunn, S.~H. Yeganeh, and V.~Jacobson, ``{BBR}: Congestion-based congestion control: Measuring bottleneck bandwidth and round-trip propagation time,'' \emph{Queue}, vol.~14, no.~5, pp. 20--53, 2016.

\bibitem{bbrv3-ietf-120-slides}
\BIBentryALTinterwordspacing
N.~Cardwell, Y.~Cheng, C.~S. Gunn \emph{et~al.}, ``{BBRv3: CCWG Internet-Draft Update},'' Internet Engineering Task Force (IETF), Tech. Rep., July 2024, presented at IETF 120. [Online]. Available: \url{https://datatracker.ietf.org/meeting/120/materials/slides-120-ccwg-bbrv3-ccwg-internet-draft-update-00}
\BIBentrySTDinterwordspacing

\bibitem{bbrrepo}
Google, ``{BBR - Source code},'' \url{https://github.com/google/bbr}, 2024.

\bibitem{bbr-discussion-google-groups}
B.~D. Group, ``{TCP BBR default in Ubuntu 24.04 (Linux 6.8) possibility?}'' \url{https://groups.google.com/g/bbr-dev/c/i-sZpfwPx-I/m/\_u7A-IijAgAJ}, 2024, accessed: 2024-10-07.

\bibitem{briscoe-iccrg-prague-congestion-control-04}
\BIBentryALTinterwordspacing
K.~D. Schepper, O.~Tilmans, B.~Briscoe, and V.~Goel, ``{Prague Congestion Control},'' Internet Engineering Task Force, Internet-Draft draft-briscoe-iccrg-prague-congestion-control-04, July 2024, work in Progress. [Online]. Available: \url{https://datatracker.ietf.org/doc/draft-briscoe-iccrg-prague-congestion-control/04/}
\BIBentrySTDinterwordspacing

\bibitem{cubic-vs-reno}
S.~Chavan, ``{Should paced TCP Reno replace CUBIC in Linux?}'' in \emph{2016 8th International Conference on Communication Systems and Networks (COMSNETS)}, 2016, pp. 1--8.

\bibitem{bbr-2019-ware}
\BIBentryALTinterwordspacing
R.~Ware, M.~K. Mukerjee, S.~Seshan, and J.~Sherry, ``{Modeling BBR's Interactions with Loss-Based Congestion Control},'' in \emph{{Proceedings of the Internet Measurement Conference}}, ser. IMC '19.\hskip 1em plus 0.5em minus 0.4em\relax New York, NY, USA: Association for Computing Machinery, 2019, p. 137–143. [Online]. Available: \url{https://doi.org/10.1145/3355369.3355604}
\BIBentrySTDinterwordspacing

\bibitem{pam-bbr3}
\BIBentryALTinterwordspacing
D.~Zeynali, E.~N. Weyulu, S.~Fathalli, B.~Chandrasekaran, and A.~Feldmann, ``{Promises and Potential of BBRv3},'' in \emph{Passive and Active Measurement: 25th International Conference, PAM 2024, Virtual Event, March 11–13, 2024, Proceedings, Part II}.\hskip 1em plus 0.5em minus 0.4em\relax Berlin, Heidelberg: Springer-Verlag, 2024, p. 249–272. [Online]. Available: \url{https://doi.org/10.1007/978-3-031-56252-5\_12}
\BIBentrySTDinterwordspacing

\bibitem{bbr2-fluid-model-imc-22}
\BIBentryALTinterwordspacing
S.~Scherrer, M.~Legner, A.~Perrig, and S.~Schmid, ``{Model-based insights on the performance, fairness, and stability of BBR},'' in \emph{Proceedings of the 22nd ACM Internet Measurement Conference}, ser. IMC '22.\hskip 1em plus 0.5em minus 0.4em\relax New York, NY, USA: Association for Computing Machinery, 2022, p. 519–537. [Online]. Available: \url{https://doi.org/10.1145/3517745.3561420}
\BIBentrySTDinterwordspacing

\bibitem{our_artifacts_github_repo}
F.~B. Sarpkaya and F.~Fund, ``{L4S-PAM2025},'' \url{https://github.com/fatihsarpkaya/L4S-PAM2025}, 2024.

\end{thebibliography}

\end{document}